\documentclass[twocolumn,showpacs,preprintnumbers,amsmath,amssymb]{revtex4}
\usepackage{graphicx}
\usepackage{dcolumn}
\usepackage{bm}

\begin{document}

\title{Strong disorder renormalization group on fractal lattices: \\
Heisenberg models and magnetoresistive effects in tight binding models}
\author{R. M\'elin}
\affiliation{
Centre de Recherches sur les Tr\'es Basses
Temp\'eratures\footnote{U.P.R. 5001 du CNRS, Laboratoire
conventionn\'e avec l'Universit\'e Joseph Fourier},
B. P. 166,
F-38042 Grenoble, France}

\author{B. Dou\c{c}ot}
\affiliation{
Laboratoire de Physique Th\'{e}orique et Hautes
\'{E}nergies, CNRS UMR 7589,  \\
Universit\'{e} Paris VI and VII,
4 place Jussieu, 75252 Paris Cedex 05, France}
\author{F. Igl\'oi}
\affiliation{
Research Institute for Solid State Physics and Optics, 
H-1525 Budapest, P.O.Box 49, Hungary}
\affiliation{
Institute of Theoretical Physics,
Szeged University, H-6720 Szeged, Hungary}

\begin{abstract}

We use a numerical implementation of the strong disorder
renormalization group (RG) method to study the
low-energy fixed points of random Heisenberg and tight-binding models on
different types of fractal lattices.
For the Heisenberg model new types of infinite disorder
and strong disorder fixed points are found. 
For the tight-binding model we add
an orbital magnetic field and use both diagonal and off-diagonal disorder.
For this model besides the gap spectra we study also the fraction of frozen
sites,
the correlation function, the persistent current and the two-terminal current.
The lattices with an even number of sites around each elementary plaquette
show a dominant $\phi_0=h/e$ periodicity. 
The lattices with an odd number
of sites around each elementary plaquette show a dominant $\phi_0/2$
periodicity at vanishing diagonal disorder,
with a positive weak localization-like
magnetoconductance at infinite disorder fixed points. 
The magnetoconductance with
both diagonal and off-diagonal disorder depends on the symmetry of the
distribution of on-site energies.
\end{abstract}

\pacs{Valid PACS appear here}
\maketitle

\section{Introduction}

The interplay between quantum fluctuations, correlations and quenched disorder
is a notoriously difficult problem of solid state theory. Recently, however,
a strong disorder renormalization
group (RG) approach was introduced 
for a class a low-dimensional random quantum systems, for a review see\cite{review}.
The idea of this method has been introduced by Ma and Dasgupta\cite{MaDasgupta} and has been
widely used after Fisher had shown\cite{fisher,fisherxx} that the method
provides asymptotically exact results for some one-dimensional
problems. 

The applicability of the RG method is due to the nature of the low-energy fixed points in these
systems. In the so called infinite disorder fixed points at larger scales the disorder grows
without limits and therefore completely dominates the low-energy, low-temperature properties
of the systems. In an infinite disorder fixed point both dynamical and static singularities
are expected, described exactly by the RG method at low energy. 
The so called strong disorder fixed points, in which the disorder has a
limited strength,
occur in different systems. At
these fixed points only the dynamical singularities are expected to be calculated
asymptotically exactly. Infinite disorder fixed points can be found in several
1D systems,
for instance
in the random transverse field Ising chain\cite{fisher}, random antiferromagnetic
Heisenberg chain\cite{fisherxx}, tight-binding model, equivalent in 1D to
the random XX model by a Jordan-Wigner transformation.
Strong disorder fixed points are
characterized by Griffiths phases\cite{griffiths,igloi99} in the Heisenberg chain with randomly
mixed ferro- and antiferromagnetic couplings\cite{westerberg} and in random quantum ladders\cite{ladders}.

In higher dimensional systems we have less information about the properties of
low-energy fixed points. Models with discrete symmetry, such as the random quantum Ising
model, generally have an infinite disorder fixed point\cite{ising2D}, whereas models with
continuous symmetry, such as the random Heisenberg model have a strong disorder fixed point\cite{heis2D}.
For the random tight-binding model on the square lattice a logarithmically infinite disorder
fixed point is conjectured\cite{gade,huse}. Generally, in higher dimensional systems the RG method
can be implemented numerically and the analysis of the results is often difficult due
to finite size effects.

The core subject of our article is devoted to investigating the properties of
the random tight-binding model with diagonal and off-diagonal disorder on
fractal lattices through a strong disorder RG.
We also make a comparison to random networks of conductors
on Euclidean geometries treated by Green's functions, and
to make contact with previous works on spin model, we discuss also the random
Heisenberg model on fractal lattices treated by a strong disorder RG.
The Heisenberg and tight-binding models have different
low-energy fixed points in 1D and in 2D and we want to investigate the properties
of this dimensional cross-over using different types of fractal lattices. We
use fractals with a sparse structure, so that the RG converges much more rapidly
than on the compact higher dimensional Euclidean lattices. Besides the
Sierpinski gasket for which we find strong disorder ({\it i.e.} not infinite disorder) fixed points, we study a
family of quasi-1D lattices that can be cut into two pieces of arbitrary sizes
by disconnecting only two bonds, for which we find 
infinite disorder fixed points. We also investigate the permanent
current and the two-terminal conductance. We give a general argument
showing that the permanent current is $\phi_0/2$-periodic for
a tight-binding model with an odd number of sites around
each elementary plaquette, with a symmetric diagonal disorder,
and with a chemical potential in the middle of the band.
We obtain an agreement between this general argument, the permanent
current calculated from the RG, and the two-terminal current
calculated from the RG and from Green's functions on Euclidean lattices.

The structure of the paper is the following. The models and the fractals are introduced
in Sec.\ref{models} and the basic ingredients of the strong disorder RG method are
presented in Sec.\ref{RG}. Our results are given in Sec.\ref{sec:heis} and \ref{sec:tb-max},
for the Heisenberg model and for the tight-binding model, respectively.
We close our paper
by a Conclusion. Some details are provided in the Appendices.

\section{Models and lattices}
\label{models}
\subsection{Random Heisenberg model}
The random Heisenberg Hamiltonian is given by
\begin{equation}
{\cal H}=\sum_{\langle k,l \rangle}
J_{k,l} {\bf S}_k . {\bf S}_l\;,
\label{hamilton_h}
\end{equation}
where the sum runs over all neighboring pairs of sites, and
where ${\bf S}_k$ corresponds to a spin-1/2.
In the antiferromagnetic model we use
$0<J_{k,l}<J_{\rm max}$ and the exchange distribution
\begin{equation}
{\cal P}(J) =
{\cal F}_{\alpha}(J,J_{\rm max}) \equiv \alpha
\frac{J_{\rm max}^{-\alpha}}{J^{1-\alpha}}\;.
\label{alpha}
\end{equation}
For the Heisenberg model with a symmetric distribution 
of exchanges we use
$-J_{\rm max}<J_{k,l}<J_{\rm max}$.
The absolute value of $J$ is distributed according to
Eq.~(\ref{alpha}).
The motivation for introducing singular exchange distributions 
is that the strong disorder RG generates by itself singular
renormalized coupling
distributions even if we start from the regular
distribution with $\alpha=1$.

\subsection{Random tight-binding model}

The random tight-binding 
Hamiltonian contains both off-diagonal disorder (terms
$t_{k,l}$) and diagonal disorder (terms $\epsilon_k$):
\begin{equation}
{\cal H}=\sum_{\langle k,l \rangle} \left[ t_{k,l}(\phi)
c_l^+ c_k + t_{l,k}(\phi) c_k^+ c_l \right]
- \sum_k \epsilon_k c_k^+ c_k\;.
\label{hamilton_tb}
\end{equation}
The hopping integrals $t_{k,l}(\phi)$ are complex numbers
and depend on the value of the magnetic flux $\phi$
through an elementary plaquette:
\begin{equation}
\label{eq:t-def}
t_{k,l}(\phi)=t_{k,l} \exp{\left(
\frac{2 i \pi}{\phi_0}
\int_{{\bf r}_k}^{{\bf r}_l}
{\bf A}.d {\bf r}
\right)}
,
\end{equation}
where the integral over the vector potential is from
${\bf r}_k$ to ${\bf r}_l$, the two points corresponding to
the lattice sites with labels $k$ and $l$.
The variables
$t_{k,l}$ correspond to the hopping integrals
in the absence of a magnetic field.

The on-site energies $\epsilon_k$ are chosen in
the distribution
${\cal P}(\epsilon)={\cal F}_{\alpha_\epsilon}
(\epsilon,\epsilon_{\rm max})$, where ${\cal F}$ is
given by Eq.~(\ref{alpha}), and the
hopping variables $t_{k,l}$ are chosen in the distribution
$
{\cal P}(t) ={\cal F}_{\alpha_t}(t,t_{\rm max})$.
To avoid multiplying the number of
parameters we restrict to the case $\alpha_\epsilon=\alpha_t$
and use the notation $\alpha=\alpha_\epsilon=\alpha_t$.

\subsubsection{Correlation functions}
\label{eq:def-corre}

The {\it correlation functions} of the tight-binding models are defined by
\begin{eqnarray}
\label{eq:C1-def}
{\cal C}_1(R) &=& 
\frac{\lambda}{{\cal N}(R)}
\overline{\sum_{\langle k,l \rangle}
\langle c^+_k c_l \rangle}\\
\label{eq:C2-def}
{\cal C}_2(R) &=&
\frac{\lambda}{{\cal N}(R)}
\overline{\sum_{\langle k,l \rangle}
\langle c^+_k c_l \rangle^2}
,
\end{eqnarray}
with the constraint that the distance between sites
$k$ and $l$ is equal to $R$. Here and in the following
$\langle \dots \rangle$ stands for the thermal average
(corresponding to the average in the ground state
at $T=0$), and
the overline denotes averaging over quenched disorder.
${\cal N}(R)$ is the number of sites at distance $R$ and
$\lambda$ is a normalization constant.

We first choose
$\lambda$ in such a way that the histograms
are normalized to unity:
\begin{equation}
\sum_{R \ne 0} C_{1,2}(R) = 1
.
\end{equation}
The other possibility is to use
unnormalized histograms corresponding to $\lambda=1$,
that will be noted ${\cal C}_{1,2}'(R)$.
${\cal C}_2'(R)$ will be used for comparing
the conductance at different magnetic fields.

\subsubsection{Conductance, current and persistent current}

The {\it differential conductance} ${\cal G}(V,T,R)$
at a fixed temperature $T$, voltage $V$
and distance $R$ is defined
as the derivative of the two-terminal current
${\cal I}(V,T)$
with respect to the applied voltage:
\begin{equation}
{\cal G}(V,T,R)= \frac {d}{d V}
{\cal I}(V,T)
\label{conduct}
.
\end{equation}
For simplicity
we carry out an averaging over all pairs of
sites at distance $R$ on the lattice, so that we
cannot discuss the geometrical effects related to the position
at which an extended contact has been connected to the
fractal lattice.

The {\it persistent current} $J(\phi,T)$
circulating in the system in the absence of an applied voltage
is equal to the derivative
of the energy $U(\phi,T)$ with respect to the magnetic flux:
\begin{equation}
\label{eq:J-orb}
J(\phi,T) = -
\frac{\partial}{\partial \phi} U(\phi,T)
.
\end{equation}
\subsection{The fractal lattices}
\begin{figure}
\includegraphics [width=1. \linewidth]{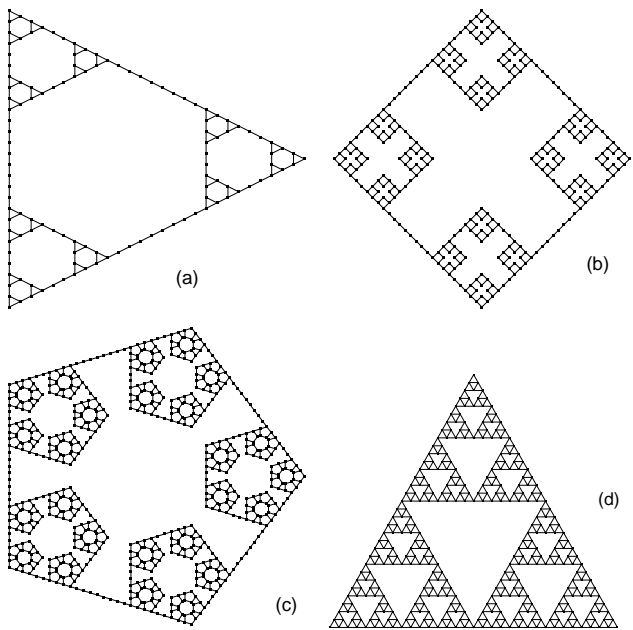}
\caption{Representation of the fractal lattices
with $n=3$, $g=4$ (a); $n=4$, $g=4$ (b);
$n=5$, $g=5$ (c); and the Sierpinski gasket
with $g=6$ (d)
\label{fig:dessin}
}
\end{figure}
We use a family of fractal lattices based on
regular polygons with
$n$ sides. We denote by $g$ the number of generations.
The lattices with $n=3,4,5$ are shown on
Fig.~\ref{fig:dessin}-(a), (b) and (c). 
In the case $n=3$ the number of sites $N_s$ and the number
of links $N_l$ is
\begin{eqnarray}
N_s &=& \left[ g+\frac{1}{2} \right] 3^{g-1} + \frac{3}{2}\\
N_l &=& 3^g 
.
\end{eqnarray}
In the case $n>3$ we define $A=(n^2-4 n+5)/[(n-1)(n-3)]$,
$B=-n/[3(n-3)]$, $C=n/(n-1)$, $A'=(n-2)/(n-3)$.
We have
\begin{eqnarray}
N_s &=& A n^g + 3^g B + C \\
N_l &=& A' n^g + 3^g B
.
\end{eqnarray}
The linear length is equal to $3^g$ so that the fractal
dimension is $d_f=\ln{n}/\ln{3}$.

We also use the standard Sierpinski gasket shown on
Fig.~\ref{fig:dessin}-(d).
The number of sites and links are given by
\begin{eqnarray}
N_s &=& 3+\frac{3}{2} \left[ 3^{g-1}-1 \right]\\
N_l &=& 3^g 
.
\end{eqnarray}
The linear length is given by $L=2^g$ so that the fractal
dimension is $d_f=\ln{3}/\ln{2}$.

\section{Strong disorder RG approach}
\label{RG}
The strong disorder RG approach\cite{review} as initiated by Ma and
Dasgupta\cite{MaDasgupta} works in the energy space and at each step a term in
the Hamiltonian is eliminated to which the largest excitation energy (gap),
denoted by $\Omega$, is associated. During this decimation process the
energy scale
$\Omega$ is lowered, and the  number of sites
of the lattice $n(\Omega)$, divided by the initial
number of sites, is reduced. At the same time new couplings are generated
between the remaining sites which were originally connected
to the decimated sites.

\subsection{RG rules}

\subsubsection{Heisenberg model}
During renormalization the Heisenberg model, if the lattice is more complex than the
linear chain, it
will transform into a set of spins of different size between
them if there
are both ferromagnetic and antiferromagnetic couplings\cite{review}.
The renormalization rules, as given in the
literature\cite{westerberg,melin00,ladders,heis2D}
involve two processes: spin-cluster formation and singlet formation.
Two spins $S_1$ and $S_2$ coupled by a strong exchange $|J_{1,2}| \sim \Omega$
give rise to a spin $S=S_1+S_2$ if $J_{1,2}<0$
(ferromagnetic coupling) and to a spin
$S=|S_1-S_2|$ if $J_{1,2}>0$ (antiferromagnetic coupling) and $S_1 \ne S_2$.
During this spin-cluster formation process the renormalized couplings are
calculated  by first order perturbation theory.
There is singlet formation if
$S_1=S_2$ are coupled antiferromagnetically, in which case
the pair of spin $S_1$ and $S_2$ is eliminated
and the renormalized couplings are
calculated by second order perturbation theory. The 
expression of the renormalized couplings can be found in
the literature\cite{westerberg,melin00,ladders,heis2D}.

\subsubsection{Tight-binding model}

In the case of the tight-binding model there are two
transformations. (i) If the largest gap of the system is
associated to a diagonal term, say $\epsilon_k=\Omega$, a fermion or a hole is frozen
at site $k$, and this site is decimated out. We will refer to this process
as an $\epsilon$-transformation.
(ii) If the largest gap of the system is
associated to an off-diagonal term, say $t_{k,l}=\Omega$, a fermion is frozen in a ``dimer''
made of sites $k$ and $l$ and this dimer is decimated out. We will refer to
this process as a
$t$-transformation. The RG transformations for a bipartite lattice with $\epsilon_k=0$, for all $k$
involves only the $t$-transformation and can be
found already in the literature\cite{huse}. A specificity of the
non bipartite lattice RG transformations 
given in Appendix~\ref{app:RG} is that diagonal disorder
can be self-generated by the RG, even though
diagonal disorder is not present in the initial condition. 

\subsection{RG flow}
\label{flow}
During renormalization we monitor the distribution function of the different
parameters of the Hamiltonians in Eqs. (\ref{hamilton_h}) and (\ref{hamilton_tb})
as a function of the energy scale $\Omega$. In different systems the
distribution of couplings,
say $P(J,\Omega)$, might have different
characteristics as the fixed point, $\Omega^*=0$, is approached.

\subsubsection{Infinite disorder fixed point}
\label{inf_dis}
 At the so called
infinite disorder fixed points the distributions broadens without limits, so that
the ratio of typical renormalized couplings goes to zero or to infinity. In this
fixed point it is convenient to introduce the
log-energy scale, $\Gamma=-\ln(\Omega_0/\Omega)$ (where
$\Omega_0$ is a reference energy), related to the length-scale,
$L \sim n(\Gamma)^{-1/d}$, as
\begin{equation}
\Gamma \sim L^{\psi}\;,
\label{Gamma_L}
\end{equation}
where $n(\Gamma)$ is the density of non-decimated points.
The exponent $\psi$ is one of the exponents
characterizing the universality class of the infinite
disorder fixed point.

This type of fixed point is realized, among others,
for the random antiferromagnetic Heisenberg spin chain and for the random
tight-binding model in 1D, for which $\psi=1/2$ in the two cases.
In an infinite disorder fixed point the distribution of the lowest gap
$\Delta$ follows the
functional form:
\begin{equation}
P_L(\Delta){\rm d} \Delta= L^{-\psi}\tilde{P}\left[\frac{\ln{\Delta}}{L^\psi}\right] {\rm d} \ln \Delta\;.
\label{eq:scaling-psi}
\end{equation}

\subsubsection{Logarithmically infinite disorder fixed point}

At this fixed point the relation between log-energy scale and length-scale is:
\begin{equation}
\Gamma \sim (\ln L)^{\omega+1}\;,
\end{equation}
thus the exponent $\psi$ is formally zero.
This type of behavior is found in the random tight-binding model on the square lattice. 
Gade\cite{gade} found $\omega=1$, whereas Motrunich {\it et al.}\cite{huse}
conjectured $\omega=1/2$, see also Ref.\cite{mudry}. At this fixed point the distribution of the lowest gap is given by:
\begin{equation}
P_L(\Delta){\rm d} \Delta= (\ln L)^{\omega+1}
\tilde{P}\left[\frac{\ln{\Delta}}{(\ln L)^{-\omega-1}}\right] {\rm d} \ln \Delta\;.
\label{eq:scaling-om}
\end{equation}

\subsubsection{Strong disorder fixed point}
\label{strong_dis}
At this fixed point the relation between energy scale and length-scale is:
\begin{equation}
\Omega \sim L^{-z}\;.
\end{equation}
where $z$ is the disorder induced dynamical exponent.
To obtain the true dynamical
exponent $z_0$ of the system, one should also consider deterministic quantum fluctuations which
could results in a vanishing gap and thus a quantum dynamical exponent $z_q$. In the non-random
system $z_0=z_q$, whereas in the presence of quenched disorder $z_0={\rm max}(z_q,z)$. Thus, for
$z>z_q$ the dynamical singularities of the system are determined by disorder effects, otherwise
disorder will results in confluent (subdominant) singularities. (In the infinite and logarithmically
infinite disorder fixed points $z$ is formally infinite.)

Strong disorder fixed points are found, among others in random antiferromagnetic Heisenberg models
on the square and cubic lattices\cite{heis2D} and generally this behavior is
related to quantum Griffiths phases\cite{griffiths,igloi99,review}. 

At this fixed point the distribution of the lowest gap is given by:
\begin{equation}
P_L(\Delta){\rm d} \Delta= L^{-z}\tilde{P}\left[\frac{\Delta}{L^z}\right] {\rm d} \Delta\;.
\label{eq:scaling-conv}
\end{equation}
If the excitations are localized, then for a small, fixed $\Delta$, $P_L(\Delta)$ is
proportional to the volume, {\it i.e.} $P_L(\Delta) \sim L^d$,
from which follows in the limit large $L$ limit: 
\begin{equation}
P(\Delta) \sim \Delta^{-1+d/z}\;.
\label{eq:P_Delta}
\end{equation}

This relation is used in numerical calculations to determine the dynamical exponent.
The asymptotic distribution of the gaps at a strong disorder fixed point has
a power-law form, that should be compared to the initial distribution in
Eq.(\ref{alpha}). There are random quantum systems in which $z$ depends on the
initial disorder and $z$
is approximately proportional to $1/\alpha$. This can be found,
among others for random Heisenberg spin ladders\cite{ladders}. In other systems with
frustration, $z$ is disorder independent (see the
three-dimensional spin glass models\cite{heis2D}), and large spins
are formed in these cases, as discussed in the following section.

\subsubsection{Singlet and large spin formation}

In random Heisenberg models during decimation the typical size of the generated new spins
can be of two types. i) The renormalized spins are typically $S=1/2$ (or $O(1)$) and do not
grow under renormalization. In this case in the ground state typically singlets are formed
and we refer to this process as random singlet formation. This type of behavior is found
in the 1D random antiferromagnetic Heisenberg model\cite{fisherxx}. ii) Another possibility
is that during renormalization large spins are formed, the size of which continuously
grows during renormalization. The
typical size of the effective spin, $S_{eff}$, is asymptotically related to the number of
decimated spins, $N$ as:
\begin{equation}
S_{eff} \sim N^{\zeta}\;.
\label{zeta}
\end{equation}
The characteristic exponent $\zeta$ would be  $\zeta=1/2$ in an uncorrelated decimation
approximation\cite{westerberg,heis2D}. Large spin formation can be found,
among others in the $\pm J$ Heisenberg models in one-\cite{westerberg},
two-\cite{heis2D}, and three-dimensions\cite{heis2D}.

\subsection{RG calculation of correlations and transport in the random tight binding model}

\subsubsection{Correlation functions}
\label{sec:calc-C}
During renormalization, as $\Omega$ is gradually decreased,
some electrons and holes becomes localized
at given sites and do not contribute to the correlation functions
in Eq.(\ref{eq:C1-def}) and (\ref{eq:C2-def}). There is, however, a
complementary process by
which electrons are bound on delocalized orbitals connecting remote sites, say $k$ and $l$.
Therefore correlations between $k$ and $l$ are $1/2$, if $\tilde{t}_{kl}$ at the
decimation energy, $|\tilde{t}_{kl}|=\Omega$, is positive and it is $-1/2$ otherwise. These
{\it rare regions} dominate the average correlation function.
{\it Typical correlations}, measured between two sites that are not involved in a
delocalized bound state, are much weaker and scale as $\sim \tilde{t}_{kl}/\Omega$. This is indeed
negligible if the distribution of $\tilde{t}_{kl}$ is broad. Thus we obtain for the
leading behavior of the average correlation function,
\begin{equation}
{\cal C}_1(\Omega,R)=
\frac{\lambda}{2{\cal N}(R)}
\sum_{\langle k,l \rangle}
s_{kl}\;,
\label{eq:C1-av}
\end{equation}
where the sum is over all frozen pairs at
 distance $R$, and $s_{kl}={\rm sign}( \tilde{t}_{kl})$.
We have a similar expression also for ${\cal C}_2(\Omega,R)$.

\subsubsection{Conductance and current}

Here we consider the system with a small applied voltage $V$ and at a small temperature $T$.
This will generate an energy scale in the system, $\Omega(V,T)={\max} (eV,T)$, and the renormalization
process stops at $\Omega=\Omega(V,T)$.
The average value of the differential conductance
is dominated by rare regions, in which, at
energy scale $\Omega(V,T)$ a frozen extended
electron state connects the two electrodes, that are separated by a distance $R$. Since
the same rare regions are dominant for the average unnormalized correlation function ${\cal C}_2'(\Gamma,R)$
we obtain the relation:
\begin{equation}
{\cal G}(V,T,R) = \frac{e^2}{h} {\cal C}'_2(\Omega(V,T),R)\;.
\label{eq:tr-form}
\end{equation}
A detailed derivation of Eq.(\ref{eq:tr-form}), based on the
perturbative theory of transport processes
can be found in Appendix~\ref{app:tr}. 

The total current ${\cal I}_{\rm tot}(V,T,R)$ in Eq.(\ref{conduct}) is obtained by integrating
the differential conductance (\ref{eq:tr-form}) over voltage, which for a small fixed
temperature, $T \ll eV$, is equivalent to an integration over the energy scale, $\Omega$:
\begin{equation}
{\cal I}_{\rm tot}(V,T,R)
\propto
\int_0^{\Omega(V,T)} {\cal G}(\Omega',R)  d \Omega'\;.
\label{eq:current-tot}
\end{equation}
Eq.~(\ref{eq:current-tot}) constitutes a generalization of
the Kubo formula since
${\cal I}_{\rm tot}(V,T,R)$ is not
necessarily linear in $V$ in the limit $V \rightarrow 0$.
The differential conductance for a large separation between the contacts
corresponds to low energies, so that we use the standard form of the RG
asymptotically exact in the low energy limit at an
infinite disorder fixed point.

\subsubsection{Persistent current}
\label{sec:persist-intro}
The zero temperature
persistent current in Eq.(\ref{eq:J-orb}) is related to the
average energy
\begin{equation}
\label{eq:U-def}
U(\phi,T)=\int_{\Omega(T)}^{\Omega_0}
\left[ -\Omega p_t(\Omega) -
\frac{\Omega}{2} p_\epsilon(\Omega) \right]
,
\end{equation}
where $p_t(\Omega)$ and $p_\epsilon(\Omega)$ are
the probabilities to perform respectively a $t$-
or and $\epsilon$-transformation at energy scale
$\Omega$. We have subtracted
in Eq.~(\ref{eq:U-def}) the energy $U(\phi,+\infty)$
at infinite temperature. The infinite temperature energy
in a $t$ transformation is vanishingly small since the
spectrum of a dimer is symmetric. The infinite temperature
energy in an $\epsilon$-transformation is equal to
$\epsilon/2$. The $\epsilon$ state is occupied only below
the Fermi level, leading to the energy $\epsilon
\theta(-\epsilon)$ (with $\theta(\epsilon)=1$ if
$\epsilon>0$ and $\theta(\epsilon)=0$ if
$\epsilon<0$). Subtracting $\epsilon/2$ yields
$-|\epsilon|/2$, leading to Eq.~(\ref{eq:U-def}).
 
\section{Random Heisenberg models}
\label{sec:heis}
\begin{figure}
\includegraphics [width=1. \linewidth]{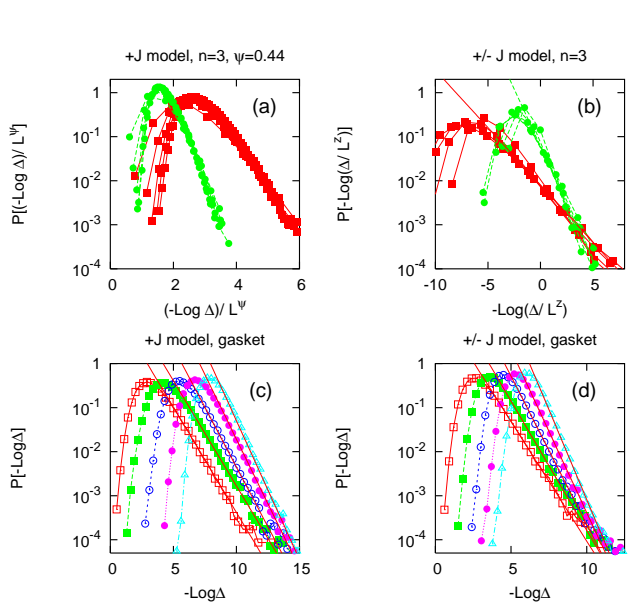}
\caption{(a) Collapse plot of the gap spectra for the antiferromagnetic
Heisenberg model on the fractal with $n=3$. (b) Collapse plot of the gap
spectra for the Heisenberg model with a symmetric distribution
of exchanges (noted $\pm J$ models on the figure)
on the fractal with $n=3$. We use
$\alpha=0.25$ ($\Box$, red) and $\alpha=0.5$ ($\bullet$, green)
on (a) and (b) respectively.
We collapsed the
gap spectra with $g=3,4,5,6,7$ generations. (a) corresponds to the
scaling (\ref{eq:scaling-psi}) with $\psi=0.44$
while (b) corresponds to the
conventional disorder scaling given by Eq.~(\ref{eq:scaling-conv}).
The corresponding values of $z$ are shown in Fig.~\ref{fig:psi-z-heis}.
The gap spectra of the antiferromagnetic
Heisenberg model and of the Heisenberg model with a symmetric
distribution of exchanges are shown on (c) and
(d) for $g=3$ ($\blacksquare$, red), $g=4$ ($\Box$, green),
$g=5$ ($\circ$, blue), $g=6$ ($\bullet$, purple), and
$g=7$ ($\triangle$, blue).
\label{fig:gap-specta-heis}
}
\end{figure}

\begin{figure}
\includegraphics [width=1. \linewidth]{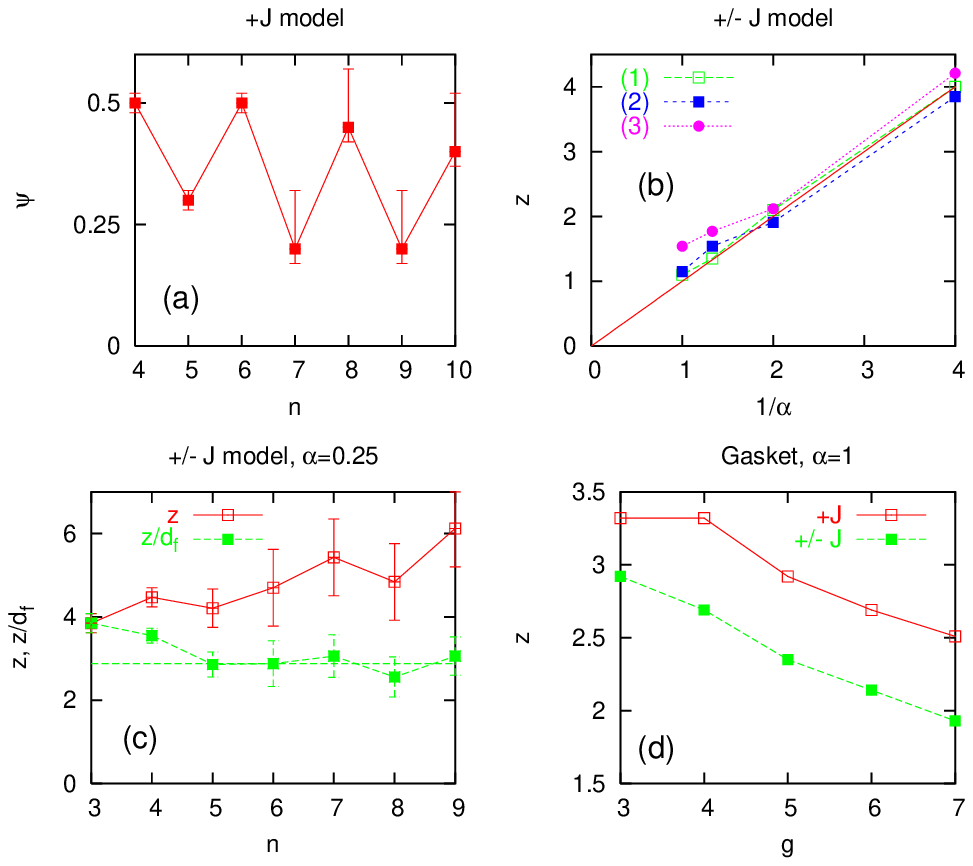}
\caption{(a) $\psi$ versus $n$ for the antiferromagnetic
Heisenberg model.
(b) $z$ versus $1/\alpha$ (proportional to the
strength of disorder) for the Heisenberg model
with a symmetric distribution of exchanges
(noted $\pm J$ model on the figure).
The curve (1) ($\Box$, green) corresponds to $n=3$ and $n=5$
with $z$ determined from the scaling form
(\ref{eq:scaling-conv}) of the gap spectra (the points for
$n=3$ and $n=5$ are superimposed). The error-bars are
comparable to the size of the symbols.
The curves (2) ($\blacksquare$, blue)
and (3) ($\bullet$, purple) correspond to
$n=3$ and $n=5$ respectively, with $z$ determined from
the slope of the rare event tail of the gap spectra.
(c) Variations of $z$ ($\Box$, red)
and $z/d_f$ ($\blacksquare$, green)
versus $n$ for Heisenberg model with a symmetric distribution
of exchanges
with $\alpha=0.25$. (d) Variation of $z$
(determined from the slope of the rare event tail
of the gap spectra)
versus $g$
for the antiferromagnetic Heisenberg model
($\Box$, red) and the Heisenberg model with a symmetric
distribution of exchanges ($\blacksquare$, green)
on the gasket with $\alpha=1$. 
\label{fig:psi-z-heis}
}
\end{figure}
To make contact
with previous works on localized spin models in 1D, 2D and 3D
we first consider
the Heisenberg model on fractal lattices
before investigating the tight-binding models on
the same lattices. In the numerical calculations we considered random
Heisenberg models on
the family of fractals with $n=3,...10$ and on the
gasket, for a purely antiferromagnetic
exchange distribution and for the
symmetric exchange distribution with mixed ferromagnetic and antiferromagnetic
couplings. The strong disorder RG procedure is implemented numerically and
the decimation is performed either up to the last remaining (large) spin or up to the
last singlet and the corresponding gap, $\Delta$, is calculated. From the distribution of
the gap $P(\Delta)$, the properties of the fixed point are determined using the
results overviewed in Sec.\ref{flow}.
The number of generations is $g=3,4,5,6,7$ for
$n=3$, $g=3,4,5,6$ for $n=4$, $g=3,4,5$ for $n=5,6$,
$g=3,4$ for $n=7,8,9,10$. We use
$\alpha=0.25, 0.5, 0.75, 1$ for $n=3,4$, and
$\alpha=0.25$ for $n=5,6,7$.
We use $250000$ realizations of disorder in each simulation.

\subsection{Distribution of the gap}

\subsubsection{Known results on regular lattices}

In 1D the random antiferromagnetic Heisenberg model\cite{fisherxx} has an infinite
disorder fixed point with the exponent $\psi=1/2$. On the contrary the $\pm J$
model in 1D has a strong disorder fixed point\cite{westerberg} with a finite dynamical
exponent, $z=2.3$ and a large-spin with $\zeta=1/2$. Quasi-one-dimensional
models, {\it i.e.} spin ladders with random antiferromagnetic couplings have a
strong disorder fixed point in which the dynamical exponent is approximately
proportional to the strength of disorder, {\it i.e.} to $\alpha^{-1}$ starting with the
distribution in Eq.(\ref{alpha})\cite{ladders}.

In 2D the random antiferromagnetic model on the square lattice, which is non-frustrated,
has a strong disorder fixed point, in which the dynamical exponent is approximately constant
for weak disorder\cite{heis2D}, but for stronger disorder $z$ becomes disorder dependent\cite{bilayer}.
On the other
hand for frustrated models (of geometric origin or for the $\pm J$ model) there is a
strong disorder fixed point with a large spin. This fixed point with $z=d$ and $\zeta=1/2$ is
disorder independent and was called a ``spin glass'' fixed point\cite{heis2D}.
There are less accurate numerical results in 3D, but it is likely that a similar scenario
as in 2D holds also in 3D\cite{heis2D}.

\subsubsection{Random antiferromagnetic model}

The results shown in Figs.~\ref{fig:gap-specta-heis}
and~\ref{fig:psi-z-heis} suggest that 
for the antiferromagnetic Heisenberg model
the fractals with $n$
even correspond to a infinite disorder fixed point
described by the usual scaling with $\psi=1/2$
whereas the fractals with $n$ odd correspond to non trivial
infinite disorder fixed points with $\psi$ lower than $1/2$
(see Fig.~\ref{fig:psi-z-heis}-(a)).
The ground state is always a singlet for $n$ even whereas there is
large spin formation for $n$ odd.
An inspection of the
RG flow indicates that spins $1$ and higher are generated
in the RG for $n$ even but correspond to an irrelevant perturbation since
they disappear soon after they are formed. The uncertainty
in the determination of $\psi$ for $n=7,8,9,10$ on
Fig.~\ref{fig:psi-z-heis} is due to finite size effects since
we cannot simulate large values of $g$ for these values of $n$.
The error-bar is estimated from the finite size effects
with $n=4$.

\subsubsection{Random $\pm J$ model}

The gap spectra of the Heisenberg model with a symmetric
distribution of exchanges
are well described by the strong disorder scaling
(\ref{eq:scaling-conv}), for both $n$ even and $n$ odd.
The value of $z$ can be extracted either from the
collapse of the gap spectra given by Eq.~(\ref{eq:scaling-conv})
or from Eq.(\ref{eq:P_Delta}) by fitting the rare event tail. We
have performed both analysis and obtained - within the error
of the calculation - identical results. The disorder induced
dynamical exponent is found proportional to the initial
strength of disorder $1/\alpha$, and we obtained
approximately:  $z\simeq 0.3 d_f / \alpha$. This corresponds to the
green dashed line in Fig.~\ref{fig:psi-z-heis}-(c) for
$\alpha=0.25$.

\subsubsection{Results on the Sierpinski gasket}

For the Sierpinski gasket we find that disorder at the end
of the RG decreases as the system size increases,
both for the antiferromagnetic Heisenberg model and the
Heisenberg model with a symmetric distribution of
exchanges (see Fig.~\ref{fig:psi-z-heis}-(d)).
The effective, size dependent dynamical exponent seems
to tend to zero as $1/\ln N$. This implies that either
i) there is a small finite gap in the system or ii) although
the gap is vanishing the dynamical correlations
decay faster than a power-law. However, due to the limited
sizes of the systems we cannot further study these scenarios.

\subsection{Distribution of the spin of the ground state}
\begin{figure}
\includegraphics [width=1. \linewidth]{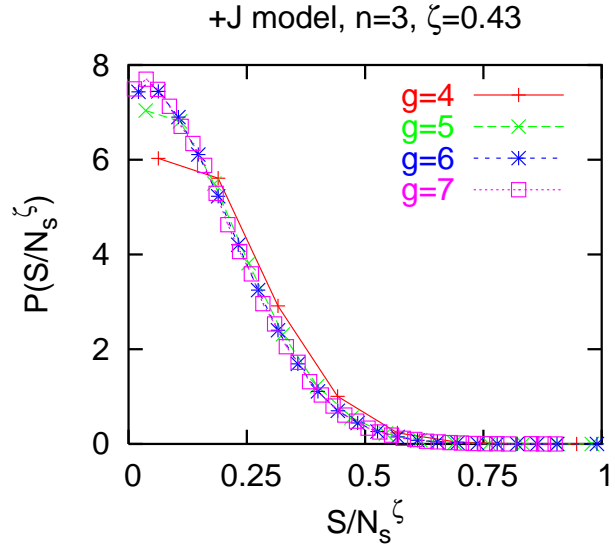}
\caption{Collapse plot of the distribution of the spin in the
ground state for the Heisenberg model on the fractal with
$n=3$, for $g=4,5,6,7$ and $\zeta=0.43$. The collapse
plot corresponds to Eq.~(\ref{zeta}).
\label{fig:PS}
}
\end{figure}
There is large spin formation for the antiferromagnetic
Heisenberg model on the fractal with $n$ odd, and for
the Heisenberg model with a symmetric distribution of exchanges
for all values of $n$. In all cases
we find that the scaling in  Eq.~(\ref{zeta}) is well
obeyed. We find for the exponent $\zeta=0.43 \pm 0.02$ for the antiferromagnetic
Heisenberg model with $n=3$ (see Fig.~\ref{fig:PS})
and $\zeta=0.5$ in all
other cases.

\subsection{Conclusion}

\begin{table*}
\caption{Low-energy fixed points of random Heisenberg models on
fractals. For a comparison results on regular 1D and 2D (square) lattices
are also presented. (AF: antiferromagnetic couplings, $\pm J$: mixed antiferromagnetic and
ferromagnetic couplings, ID: infinite disorder fixed point, SD: strong disorder
fixed point, RS: random singlet formation, LS: large spin formation.) $^{(+)}$ 
For $n=3$, $\zeta<1/2$.\label{table:1}}
 \begin{tabular}{|c|c|c|c||c|c|}  \hline
       & $n=odd$        & $n=even$      & gasket &  1D           &  2D    \\ \hline
   AF  & ID($\psi=1/2$)&ID($\psi<1/2)$ & SD($z \to 0$)(?) & ID($\psi=1/2$)&   SD \\
       & LS($\zeta=1/2^{(+)}$)& RS     &  LS($\zeta=1/2$) & RS             & RS \\ \hline
$\pm J$ & SD($z \sim 1/\alpha$) & SD($z \sim 1/\alpha$) & SD($z \to 0)$(?)& SD($z=2.3$) & SD($z=d$) \\
   & LS($\zeta=1/2$)  & LS($\zeta=1/2$)& LS($\zeta=1/2$) & LS($\zeta=1/2$) & LS($\zeta=1/2$) \\ \hline
  \end{tabular}
  \end{table*}
The empirically obtained critical behavior of the random Heisenberg model
on different fractal lattices is summarized in Table \ref{table:1}.
The low-energy fixed point seems to depend on the form (symmetry) of disorder, the
symmetry of the fractal lattice, but there is no direct relation to the
fractal dimension of the lattice. Interestingly the studied fractal lattices constitute
low-energy universality classes that can not be found in the regular 1D and
2D lattices.

\section{Random tight-binding models}
\label{sec:tb-max} 

The detail of the RG transformations of the tight-binding
model can be found in Appendix~\ref{app:RG}.
The interferences between different trajectories 
in the $t$-transformation are eliminated
if the sum of the different terms in the renormalized
coupling is replaced by their maximum,
an approximation
used in the study of spin models.
To keep interferences, we do not use the maximum in the expression
of the renormalized couplings.

\subsection{Features of the RG flow without extrinsic diagonal disorder}

\subsubsection{Fraction of frozen sites}
\label{sec:n-eps1}
\begin{figure}
\includegraphics [width=1. \linewidth]{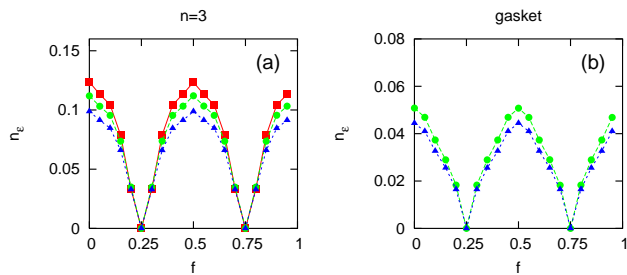}
\caption{Fraction of frozen sites $n_\epsilon$
(defined by Eq.~(\ref{eq:n-eps})) versus $f$
(a) for the fractal with $n=3$ and
$g=4$ ($\blacksquare$, red),
$g=5$ ($\bullet$, green) and $g=6$ ($\blacktriangle$, blue);
and (b) for the gasket with
with $g=4$ ($\blacksquare$, red),
$g=5$ ($\bullet$, green) and $g=6$ ($\blacktriangle$, blue) generations.
We used $\alpha=1$.
\label{fig:n-eps-intrin}
}
\end{figure}

We first consider the case where diagonal disorder is not introduced
in the initial condition. We thus start with $\epsilon_k=0$ for all $k$.
For a bipartite lattice (for instance the fractal with $n$ even)
no diagonal term is generated during the RG transformations, which
can be seen from the decimation rules given in Appendix~\ref{app:RG}.
On the contrary for a non bipartite lattice (for instance for
the fractal with $n$ odd)
also non vanishing values of $\epsilon_k$ are generated,
and at the end of the RG an average number
$N_\epsilon$ of the total number of sites $N_s$ 
has been frozen by the $\epsilon$-transformations. We have
calculated the fraction of frozen fermions,
\begin{equation}
\label{eq:n-eps}
n_\epsilon=N_\epsilon / N_s
\end{equation}
for both the fractal with $n=3$ and for the gasket.
We work in the grand canonical ensemble where the chemical
potential is fixed.

We show in Fig.~\ref{fig:n-eps-intrin} the variation of
$n_\epsilon$ as a function of $f=\phi/\phi_0$. First
we note that $n_\epsilon$ decreases slowly with increasing system size: in the
limit $N_s \rightarrow \infty$ our data indicate an
approximately logarithmic dependence, $n_\epsilon \sim 1 / \ln{N_s}$.
In a finite size system $n_\epsilon$ shows 
$\phi_0/2$-oscillations so that the number of frozen fermions vanishes at
$f=0.25, 0.75, 1.25, ...$ and there is a tendency of the system
to be more localized for $f=0, 1/2, 1 ...$, in agreement with the
positive magnetoconductance discussed later in section \ref{sec:twoter}.
The cancellation of
$n_\epsilon$ for $f=0.25,0.75,1.25,...$ can be understood from
the renormalization of $\epsilon$ in the $t$-transformation
(see Eq.~\ref{eq:epstilde-inter1}). If the sites ``1'' and ``2''
are eliminated in a three-site cluster $\{1,2,3\}$ then
the interference of the trajectories
$3\rightarrow 1 \rightarrow 2 \rightarrow 3$ 
and the trajectory
$3\rightarrow 2 \rightarrow 1 \rightarrow 3$ are such that
$\tilde{\epsilon}_3=0$, so that no diagonal
disorder is generated by the RG for these values of
$f=\phi/\phi_0$.
\subsubsection{Average coordinance}
\begin{figure}
\includegraphics [width=1. \linewidth]{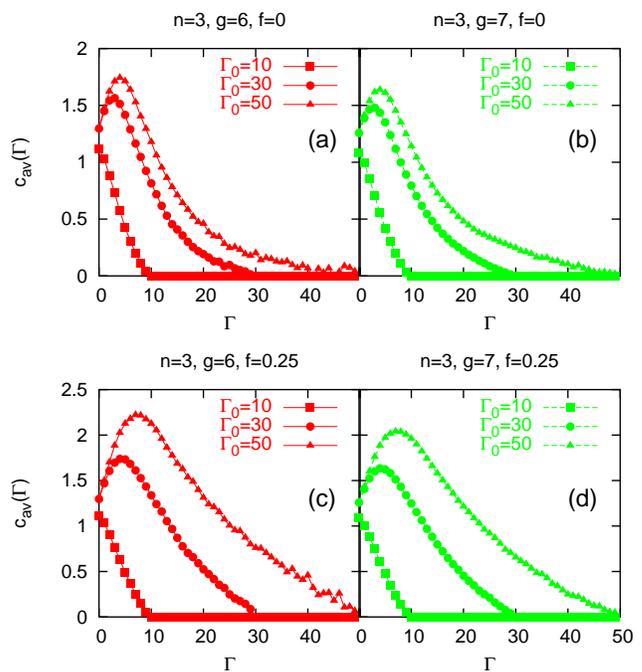}
\caption{Average coordinance of the graph of ``active'' spins
as a function of $\Gamma$, for different values of $\Gamma_0$,
for the fractal with $n=3$.
(a) corresponds to $g=6$ generations and $f=0$.
(b) corresponds to $g=7$ and $f=0$.
(c) corresponds to $g=6$ and $f=0.25$.
(d) corresponds to $g=7$ and $f=0.25$.
The different curves in each panel correspond to
$\Gamma_0=10$ ($\blacksquare$),
$\Gamma_0=30$ ($\bullet$) and
$\Gamma_0=50$ ($\blacktriangle$).
The red color is used for $g=6$ and the green color
for $g=7$. We used $\alpha=0.25$.
\label{fig:coord-n3}
}
\end{figure}

\begin{figure}
\includegraphics [width=1. \linewidth]{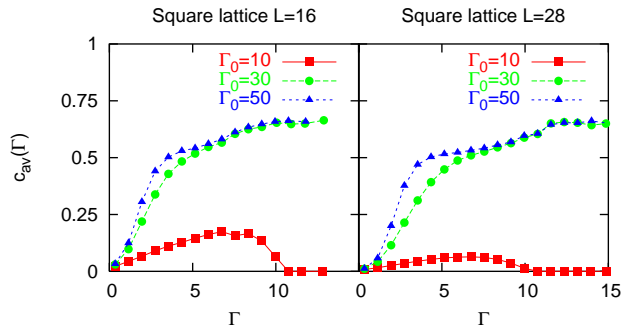}
\caption{Variation of the average clustering coefficient
$c'_{\rm av}(\Gamma)$, see text,
as a function of $\Gamma$, for different values of $\Gamma_0$,
for the Euclidian 2D square lattice with $f=0$.
(a) corresponds to $L=16$ and
(b) corresponds to $L=28$.
The different curves in each panel correspond to
$\Gamma_0=10$ ($\blacksquare$, red),
$\Gamma_0=30$ ($\bullet$, green) and
$\Gamma_0=50$ ($\blacktriangle$, blue).
We used $\alpha=0.25$.
\label{fig:2D}
}
\end{figure}

A useful information is encoded in the connectivity of the graph
of ``active'' sites (the sites that have not been eliminated 
at the energy scale $\Omega$). More precisely we consider the connectivity of
clusters of ``strongly connected'' sites, which are defined by having
hopping amplitudes $t_{k,l}$ that exceed a limiting value:
\begin{equation}
 \ln{\left[\frac{\Omega}
{|t_{k,l}|}\right]} < \Gamma_0
,
\end{equation}
where $\Gamma_0$ is a given cut-off.

We calculate the average coordinance $c_{\rm av}$
of this graph as a function of the log-energy scale $\Gamma$, where
$c_{\rm av}$ is defined as the ratio of the average
number of bonds and the average number of sites.
The average coordinance
$c_{\rm av}$ is expected to start from a value of order unity,
first increase, reach a maximum,
and then decrease to zero at large $\Gamma$ since
only a few sites are surviving at the end of the RG.

$c_{\rm av}(\Gamma)$ is shown in Fig.~\ref{fig:coord-n3}
for the fractal with $n=3$, for two sizes corresponding to
$g=6,7$, and for $f=0$ and $f=0.25$. 
It is visible that there are practically no
finite size effects for $g=6,7$. The average coordinance is
larger for $f=0.25$ than for $f=0$, compatible with the 
existence of a negative magnetoresistance discussed in
section \ref{sec:twoter}:
the system is
more connected and therefore more delocalized for $f=0.25$.

By comparison we carried out a similar
simulation for the Euclidian two-dimensional square lattice.
The finite size effects indicate that after a transient
the graph of surviving sites becomes extremely connected whereas
it is loosely connected in the case of the fractal with $n=3$.
This geometrical effect is related to the fact that the
algorithm becomes extremely slow in the case of the Euclidian
square lattice as compared to the fractal lattices.
More quantitatively it is convenient to evaluate
the clustering coefficient $c'_{\rm av}(\Gamma)$, which is
the average fraction of bonds, relative to the maximum number
of possible bonds, $N_s(N_s-1)/2$.
For sufficiently large values of $\Gamma_0$ the quantity
$c'_{\rm av}(\Gamma)$ reaches a plateau at around $0.7$,
see Fig.\ref{fig:2D}, indicating that the graph of active sites becomes strongly
connected.

\subsubsection{Gap spectra}
\label{sec:gap-spectra-tb}

\begin{figure}
\includegraphics [width=1. \linewidth]{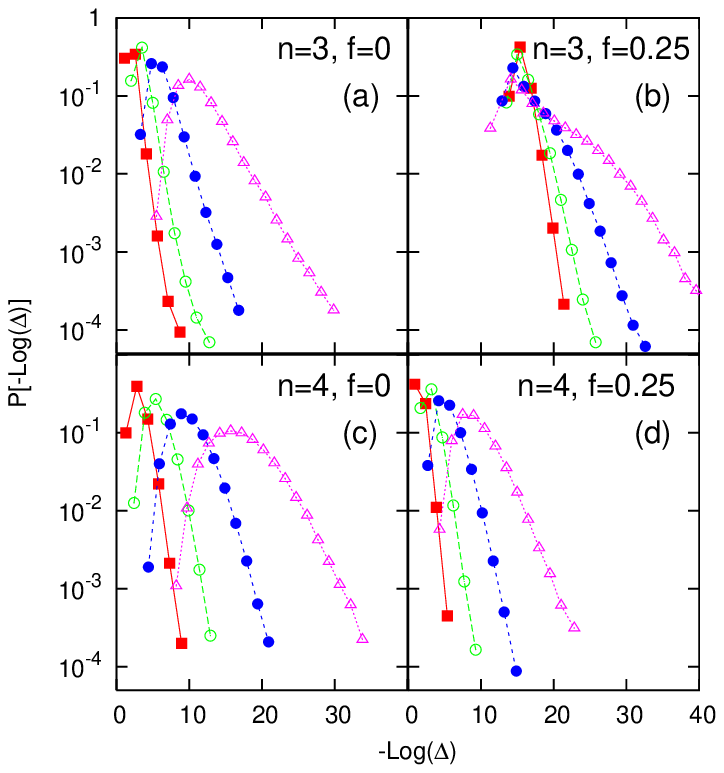}
\caption{Gap spectra of the tight binding model with 
($n=3$, $f=0$) (a), ($n=3$, $f=0.25$) (b), ($n=4$, $f=0$) (c), 
and ($n=4$, $f=0.25$) (d).
The simulations on (a) and (b) with $n=3$ correspond to
for $g=4$ ($\blacksquare$, red), $g=5$ ($\circ$, green) $g=6$
($\bullet$, blue), and $g=7$ ($\triangle$, purple).
The simulations on (c) and (d) with $n=4$ correspond to
$g=3$ ($\blacksquare$, red), $g=4$ ($\circ$, green), $g=5$
($\bullet$, blue), and $g=6$ ($\triangle$, purple).
We used $\alpha=1$.
\label{fig:gap-spectra-tb-f}
}
\end{figure}

\begin{figure}
\includegraphics [width=1. \linewidth]{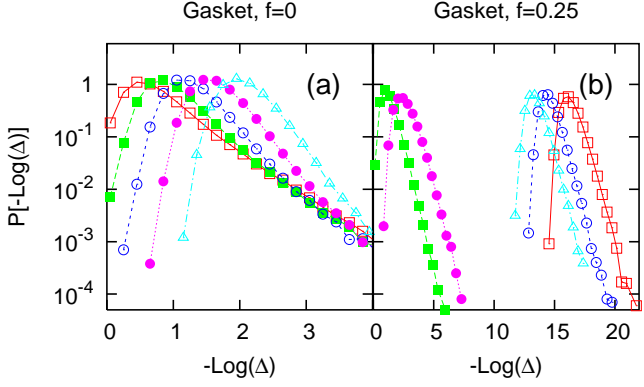}
\caption{Gap spectra of the tight binding model on the
Sierpinski gasket with $f=0$ (a) and $f=0.25$ (b)
and with $g=3$ (red, $\square$),
$g=4$ (green, $\blacksquare$),
$g=5$ (blue, $\circ$),
$g=6$ (purple, $\bullet$)
and $g=7$ (blue, $\triangle$)
 generations.
We used $\alpha=1$.
\label{fig:gap-spectra-ga}
}
\end{figure}

We show in Fig.~\ref{fig:gap-spectra-tb-f} for the fractals
with $n=3,4$ 
the gap spectra of the
random tight-binding model with $\epsilon_{\rm max}=0$
and $f=0,0.25$. In all these cases 
there is infinite disorder scaling\cite{note1}, see Sec.\ref{inf_dis}. 
The collapse plots according to Eq.~(\ref{eq:scaling-psi}) show that $\psi=0.54\pm0.04$ for ($f=0$, $n=3$),
$\psi=0.5\pm 0.03$ for ($f=0$, $n=4$),
$\psi=0.48\pm0.02$ for ($f=0.25$, $n=4$).
To obtain the scaling for ($f=0.25$, $n=3$)
we repeated the simulations in Fig.~\ref{fig:gap-spectra-tb-f}-(b)
without the last $\epsilon$-transformation and found that the scaling
is well described by $\psi=0.48 \pm 0.02$.

The gap spectra for the gasket shown in Fig.~\ref{fig:gap-spectra-ga}
indicate strong disorder scaling, see Sec.\ref{strong_dis}.
The parity effect for $f=0.25$ is similar to the fractal with $n=3$
and $f=0.25$ in Fig.~\ref{fig:gap-spectra-tb-f}, see\cite{note1}. For the 
Sierpinski gasket
the number of sites $N_s$ is odd for $g$ odd and even for
$g$ even.

\begin{figure}
\includegraphics [width=1. \linewidth]{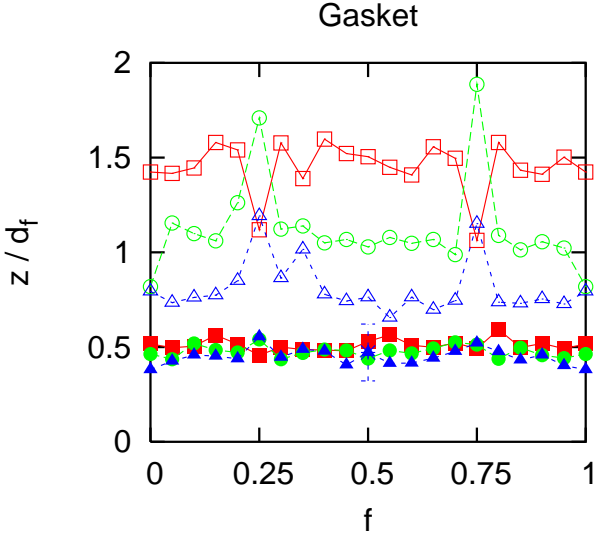}
\caption{Ratio $z/d_f$ versus $f$ for the Sierpinski gasket
with ($g=4$, $\alpha=1$) ($\blacksquare$, red),
($g=5$, $\alpha=1$) ($\bullet$, green) and
($g=6$, $\alpha=1$) ($\blacktriangle$, blue),
($g=4$,$\alpha=0.25$) ($\square$, red),
($g=5$,$\alpha=0.25$) ($\circ$, green) and
($g=6$, $\alpha=0.25$) ($\triangle$, blue).
The ratio $z/d_f$ is smaller than unity for sufficiently
large system sizes.
\label{fig:z-vs-f_ga}
}
\end{figure}

We show in Fig.~\ref{fig:z-vs-f_ga} the variation of the
ratio $z/d_f$ versus $f$ for the Sierpinski gasket, as calculated from
Eq.~(\ref{eq:P_Delta}).
For a weaker initial disorder  ($\alpha=1$) the dynamical exponent is practically
independent of $f$ and has no finite size effect, so we can estimate
$z/d_f=0.5$.
The finite size effects are also stronger 
for stronger initial disorder ($\alpha=0.25$).
The dynamical exponent for very large system sizes is probably
$f$-independent, and
the apparent variations are
within the accuracy of the numerical calculation. From the available data it is
difficult to estimate the limiting value. However, $z/d_f$ (the strength of
disorder at the fixed point) increases with $1/\alpha$ (the strength of
disorder in the initial condition), so that we might assume that the
limiting value of $z/d_f$ could be close to $0.5$ for $\alpha=0.25$.

\subsection{Normalized correlation functions}
\label{corr_funct}
\begin{figure}
\includegraphics [width=1. \linewidth]{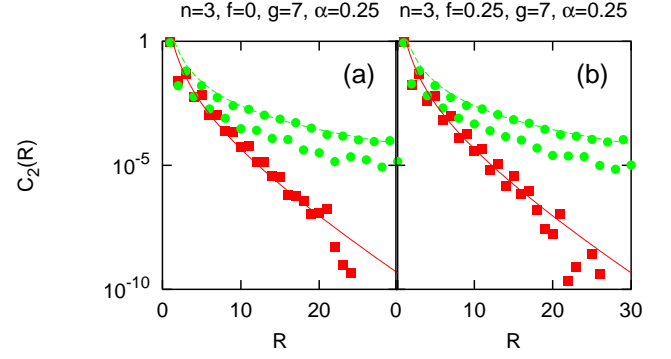}
\caption{
Normalized correlation function ${\cal C}_2(\Gamma,R)$
versus $R$ for the fractal with $n=3$ with $g=7$ generations,
${\alpha=0.25}$ and $f=0$ (a), $f=0.25$ (b).
The different curves correspond to
$\Gamma=5$ ($\blacksquare$, red),
$\Gamma=30$ ($\bullet$, green).
The solid lines correspond to the fit (\ref{eq:fit-C2}),
with $x=1.6$ and $\xi(\Gamma)=2.5$ for $\Gamma=5$,
and $\xi(\Gamma)=+ \infty$ for $\Gamma=30$.
\label{fig:C2-vs-R}
}
\end{figure}
The definition of the correlation function is given in
section~\ref{sec:calc-C}.
The correlation functions obtained from the simulations
are well fitted by 
\begin{equation}
\label{eq:fit-C2}
{\cal C}_2(\Gamma,R) \sim
\frac{1}{R^{2 x}} \exp{\left(-\frac{R}{\xi(\Gamma)}\right)}
,
\end{equation}
with a correlation length $\xi(\Gamma)$ diverging 
like $\xi(\Gamma) \sim \Gamma^{1/\psi}$ at an
infinite disorder fixed point, see
Eq.(\ref{Gamma_L}). 
For the Euclidean 1D random tight-binding model we 
have $x=1$ from the simulation,
in agreement with the analytical prediction\cite{fisherxx}.
For the fractals with $n=3$ and $n=4$,
we obtain the same values of the decay exponent $x$
and the correlation length
$\xi(\Gamma)$, for
${\cal C}_1(\Gamma,R)$ and ${\cal C}_2(\Gamma,R)$.
This is expected since the dominant contribution to the correlation
functions is due to the same type of rare regions for both functions.
The decay length of
${\cal C}_1(\Gamma,R)$ is equal to the decay length of
${\cal C}_2(\Gamma,R)$, a feature of 1D Euclidean
systems that we recover on the sparse fractal lattices.
In usual disordered systems with diagonal disorder and
no off-diagonal disorder, the decay length
of ${\cal C}_1(R)$ at zero temperature is equal to the elastic mean
free path, and the decay length of ${\cal C}_2(R)$ at zero temperature
is equal to the localization length. At finite temperature there
exists an additional exponential decay over the thermal length
$\xi_T$.
The situation considered here with
off-diagonal disorder and diagonal disorder
can be viewed as a temperature dependent mean free
path equal to the temperature dependent localization length.
Like in 1D systems,
there is no genuine diffusive regime in which the localization
length would be much larger than the elastic mean free path.
Interestingly, we still obtain a weak localization-like
magnetoresistance for the fractal with $n=3$
(see Secs.\ref{sec:dia} and \ref{sec:twoter}).

The variation and the fit of 
${\cal C}_2(\Gamma,R)$ versus $R$ is shown in Fig.~\ref{fig:C2-vs-R}
for the fractal with $n=3$. We used
$x=1.6$ for all values of $\alpha$ but, as expected,
$\xi(\Gamma)$ increases as $\alpha$ increases (not shown
in Fig.~\ref{fig:C2-vs-R}). The same value of $x$
is obtained for $n=4$ (not shown in Fig.~\ref{fig:C2-vs-R}).
We find no significative difference
in the values of $x$ and $\xi(\Gamma)$
between the cases $f=0$ and $f=0.25$.

There exists a parity effect like in 1D. Namely the correlations
between two sites at distance $R$ are much stronger if $R$ is odd.
The origin of this effect in $1D$ is that 
all
the sites in between should
be coupled in localized pairs
in order to have an extended
localized orbital between the two remote sites at distance $R$.
This is obviously only possible for $R$ odd.
Similar effect might be due to the $1D$ chains embedded in the fractal with $n=3$.

\subsection{Effect of extrinsic diagonal disorder at $f=0$}

\subsubsection{Fraction of frozen sites}
\begin{figure}
\includegraphics [width=1. \linewidth]{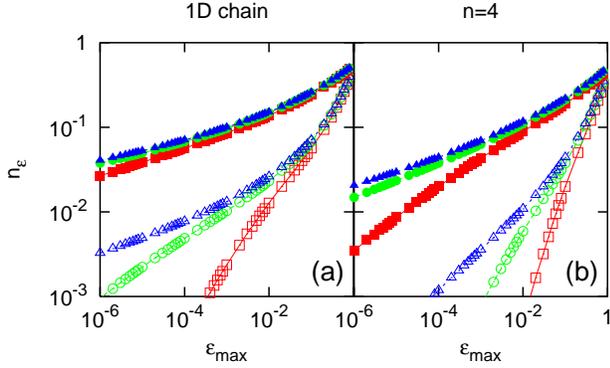}
\caption{Variation of the fraction of sites $n_\epsilon$ frozen
by an $\epsilon$-transformation as a function of
$\epsilon_{\rm max}$, the upper cut-off in the distribution
of $\epsilon$, for the 1D chain (a) and the fractal with $n=4$ (b).
For the 1D chain on (a)
we used chains of length $L=100$ ($\blacksquare$ and $\square$, red),
$L=400$ ($\bullet$ and $\circ$, green)
and $L=1600$ ($\blacktriangle$ and $\triangle$, blue).
For the fractal with $n=4$ on (b) 
we used $g=3$ ($\blacksquare$ and $\square$, red),
$g=4$ ($\bullet$ and $\circ$, green)
and $g=5$ ($\blacktriangle$ and $\triangle$, blue).
The open symbols correspond to $\alpha=1$ and
the filled symbols correspond to $\alpha=0.25$.
\label{fig:n-eps-1D-n4}
}
\end{figure}

\begin{figure}
\includegraphics [width=1. \linewidth]{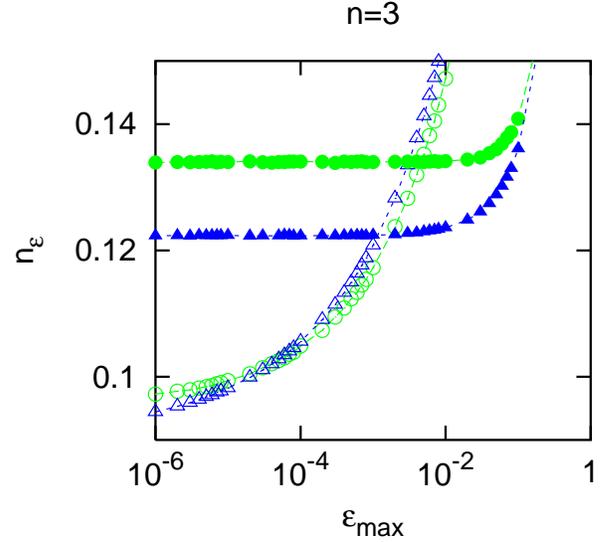}
\caption{Variation of the fraction of sites $n_\epsilon$ frozen
by an $\epsilon$-transformation as a function of
$\epsilon_{\rm max}$, the upper cut-off in the distribution
of $\epsilon$, for the fractal with $n=3$.
We used
$g=4$ ($\bullet$ and $\circ$, green)
and $g=5$ ($\blacktriangle$ and $\triangle$, blue).
The open symbols correspond to $\alpha=1$ and
the filled symbols correspond to $\alpha=0.25$.
We obtained a similar behavior for $n=5$.
\label{fig:n-eps-n3}
}
\end{figure}

We now consider the case of a finite extrinsic
diagonal disorder with $f=0$.
Increasing $\epsilon_{\rm max}$ tends to increase 
the number of $\epsilon$-transformation and thus the fraction
of frozen sites $n_\epsilon(\epsilon_{\rm max})$.
For bipartite lattices, such as the 1D
chain and the fractal with $n=3$ the limiting value of
$n_\epsilon(\epsilon_{\rm max})$ is finite in the limit
$\epsilon_{\rm max} \to 0$, even
for a finite lattice. This behavior is illustrated
in Fig.~\ref{fig:n-eps-1D-n4}. The limiting behavior of
$n_\epsilon(\epsilon_{\rm max})$
can be estimated as follows. For a small fixed $\epsilon_{\rm max}$, the RG transformation
goes along the trajectory of the $\epsilon_{\rm max}=0$ case, until the
log-energy scale
$\Gamma=\ln( \epsilon_0/\epsilon_{\rm max})$ is reached, with some reference
value $\epsilon_0$.
 At this point 
$n_\epsilon(\epsilon_{\rm max},N_s)$ is proportional to
the number of sites $N_s$.
The infinite disorder scaling given by Eq.~(\ref{Gamma_L}) corresponds to
$\ln (\epsilon_0/\epsilon_{\rm max}) \sim \Gamma \sim N_s^{\psi/d_f}$. We thus
obtain a logarithmic dependence as $n_\epsilon(\epsilon_{\rm max})
\sim (\ln (\epsilon_0/\epsilon_{\rm max}))^{d_f/\psi}$.
The numerical data in Fig.~\ref{fig:n-eps-1D-n4} are compatible with this form,
although a simple power-law form
$n_\epsilon(\epsilon_{\rm max}) \sim \epsilon_{\rm max}^y$,
with an effective exponent $y=0.14$ seems to work, too,
both for the 1D chain and the
fractal with $n=4$.

For a non-bipartite lattice, such as the fractal with $n=3$,
$n_\epsilon$ tends to a constant in the limit
$\epsilon_{\rm max} \rightarrow 0$ for a finite lattice, see Sec.\ref{sec:n-eps1}. 
This is illustrated in  Fig.~\ref{fig:n-eps-n3}.
For a small $\epsilon_{\rm max}$,
$n_\epsilon$ increases as the system size increases for bipartite (the 1D and the fractal with $n=4$)
lattices and decreases for non-bipartite (the fractal with $n=3$) lattices.

\subsubsection{Correlation length}
\label{sec:corre-length}
\begin{figure*}
\includegraphics [width=.8 \linewidth]{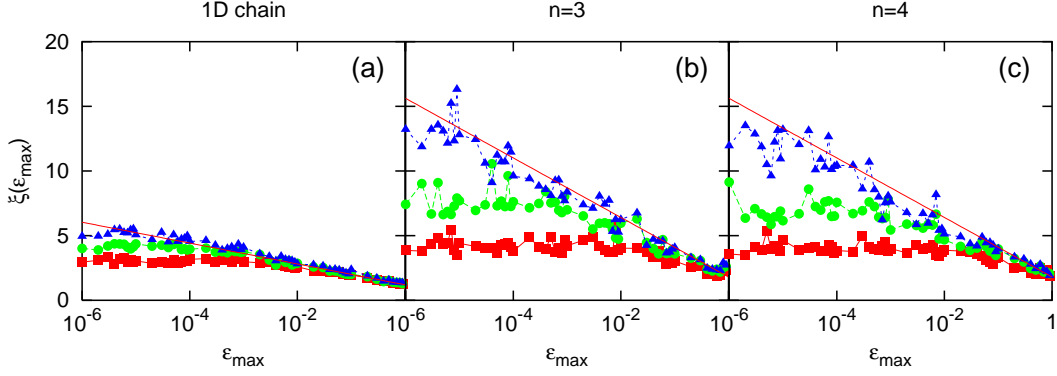}
\caption{Average correlation length $\xi_{av}(\Gamma,\epsilon_{\rm max})$ 
as a function of $\epsilon_{\rm max}$ for several values of
$\Gamma$, for the 1D chain (a), the fractal with $n=3$ (b)
and the fractal with $n=4$ (c). In (a) we use
$\Gamma=7$ ($\blacksquare$, red),
$\Gamma=9$ ($\bullet$, green),
$\Gamma=11$ ($\blacktriangle$, blue).
In (b) and (c) we use
$\Gamma=5$ ($\blacksquare$, red),
$\Gamma=8$ ($\bullet$, green),
$\Gamma=11$ ($\blacktriangle$, blue),
We use $\alpha=0.25$, $L=6400$ for the 1D chain,
$g=7$ for the fractal with $n=3$,
and $g=6$ for the fractal with $n=4$.
The fits correspond to $\xi_{\rm av}(\Gamma,\epsilon_{\rm max})
=a \ln{(\epsilon_{\rm max}/\Omega_0)}+b$,
with $a=0.35$ and $b=1.2$ for the 1D chain,
and $a=1$ and $b=1.8$ for the fractals
with $n=3$ and $n=4$.
We obtained qualitatively similar results for other
values of $\alpha$, and for the ``$\pm \epsilon$''
model with a symmetric distribution of $\epsilon$.
\label{fig:xi-eps}
}
\end{figure*}
The {\it average} correlation length $\xi_{av}(\Gamma,\epsilon_{\rm max})$ is
calculated from the average correlation function ${\cal
C}_2(\Gamma,R)$. Using the functional form in Eq.~(\ref{eq:fit-C2})
we fix the decay exponent $x$ to its value for $\epsilon_{\rm
  max}=0$ (see Sec. \ref{corr_funct}), and then extract the value of
$\xi_{av}(\Gamma,\epsilon_{\rm max})$ from the numerical data.
The numerical results are shown in Fig.\ref{fig:xi-eps} 
for the 1D chain (a) and
for the fractals with $n=3$ (b) and $n=4$ (c). In the limit 
$\Gamma \gg  \ln(\epsilon_{\rm max}/\epsilon_0)$
the average correlation length has a $\Gamma$ independent limiting value which
depends on  $\ln(\epsilon_{\rm max}/\epsilon_0)$ and the functional
form seems to be linear for all type of lattices. Here we can observe an 
analogy
with the behavior of the average correlation length in a disordered
Heisenberg chain used as a model of disordered spin-Peierls system
at a finite temperature $T$, as calculated in Ref.\cite{FM-PRB}. Making the
correspondence with the
energy scales in the two problems, $\epsilon_{\rm max} \leftrightarrow T$,
we arrive to the relation
$\xi_{av}(\Gamma,\epsilon_{\rm max}) \sim \ln(\epsilon_{\rm max}/\epsilon_0)$
for
$\Gamma \gg  \ln(\epsilon_{\rm max}/\epsilon_0)$, as observed numerically.

\begin{figure*}
\includegraphics [width=1. \linewidth]{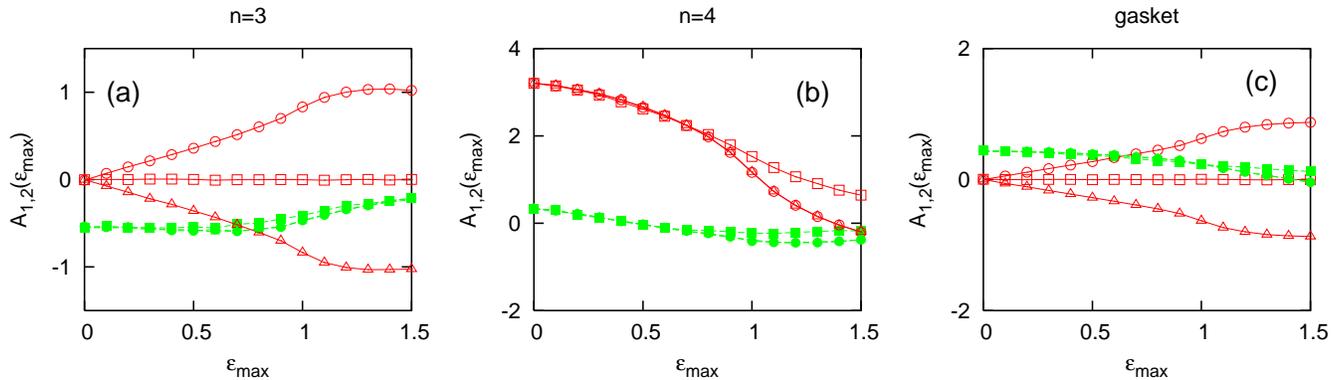}
\caption{Evolution of the harmonics $A_1(\epsilon_{\rm max})$
(open symbols,red)
and $A_2(\epsilon_{\rm max})$ (filled symbols, green)
of the permanent current,
as a function of  $\epsilon_{\rm max}$.
Panel (a), (b) and (c)
correspond respectively to
the fractals with $n=3$, $n=4$ and to the Sierpinski gasket.
The different curves within each panel correspond to
the ``$\pm \epsilon$'' model with a symmetric distribution of
$\epsilon_k$ ($\Box$ and $\blacksquare$), to
the ``$+\epsilon$'' model with $\epsilon_k>0$ ($\circ$ and
$\bullet$), and to the
``$-\epsilon$'' model with $\epsilon_k<0$ ($\triangle$
and $\blacktriangle$).
\label{fig:RG-perma}
}
\end{figure*}

\subsection{Persistent current}
\label{sec:dia}

\subsubsection{General argument}
\label{sec:general}
We expect the averaged persistent current to exhibit
a $\phi_0/2$ periodicity for any tight-binding
lattice with an odd number of sites around its elementary plaquettes,
provided the distribution of diagonal disorder is even and the
chemical potential $\mu$ is fixed equal to zero.
The following argument is inspired by the discussion given in\cite{Browne84}
for Aharonov-Bohm current oscillations in a mesoscopic ring.
The eigenvalue equation for the single-particle 
Hamiltonian~(\ref{hamilton_tb}) reads:
\begin{equation}
E\psi_{k}=-\epsilon_{k}\psi_{k}+\sum_{l}t_{kl}(\phi)\psi_{l}
\end{equation}
If we multiply this equation by $-1$, we get an eigenstate
with energy $-E$, for a tight-binding model where on-site 
potentials and hopping amplitudes have been turned into their
opposite values. On a bipartite lattice, changing 
$t_{kl}(\phi)$ into $-t_{kl}(\phi)$ is simply equivalent
to performing a gauge transformation, which leaves the
external magnetic field unchanged. But on a lattice (such as
a triangular lattice) with elementary odd cycles, this change
of sign in the hopping amplitudes is equivalent to changing the
elementary flux $\phi$ into $\phi+\phi_{0}/2$. Denoting
single-particle energy eigenvalues by $E_{\alpha}(\phi,\{\epsilon_{k}\})$,
we have then
\begin{eqnarray}
&&\sum_{\alpha}E_{\alpha}(\phi,\{\epsilon_{k}\})\theta(\mu-E_{\alpha})\\
&+&\sum_{\beta}E_{\beta}(\phi+\frac{\phi_{0}}{2},\{-\epsilon_{k}\})
\theta(E_{\beta}+\mu)=0
\end{eqnarray}
Let us then average these expressions over disorder realizations.
Since the $\epsilon_{k}$ distribution is assumed to be even, this yields:
\begin{equation}
\sum_{\alpha}\overline{E_{\alpha}(\phi)\theta(\mu-E_{\alpha})}
+\sum_{\beta}\overline{E_{\beta}(\phi+\frac{\phi_{0}}{2})\theta(E_{\beta}+\mu)}
=0
\end{equation}
The sum-rule on the single-particle spectrum yields
\begin{eqnarray}
\nonumber
&&\sum_{\beta}\overline{E_{\beta}(\phi+\frac{\phi_{0}}{2})
(\theta(E_{\beta}+\mu)+\theta(-\mu-E_{\beta}))}\\
&=&-\sum_{k}\overline{\epsilon_{k}}=0
\end{eqnarray}
So finally:
\begin{equation}
\sum_{\alpha}\overline{E_{\alpha}(\phi)\theta(\mu-E_{\alpha})}=
\sum_{\beta}\overline{E_{\beta}(\phi+\frac{\phi_{0}}{2})\theta(-\mu-E_{\beta})}
\end{equation}
For $\mu=0$, this implies the $\phi_{0}/2$ periodicity of the
system's total energy. 
The above argument also holds for the fractal
considered here with $n=3$ or for the Sierpinski gasket
even though they have also some elementary loops with an {\em even}
number of sites. This is because these larger loops enclose
an area which is an {\em even} multiple of the elementary
triangle area. So changing $\phi$ into $\phi+\phi_{0}/2$
does not change the total flux modulo $\phi_{0}$
in larger elementary loops, which is compatible with a sign
reversal of all hopping amplitudes.

A direct experimental test of these
predictions is however difficult, since we
have considered single channel tight-binding models, and it is not
possible to put a flux quantum through an area of atomic size.
The qualitative difference between bipartite
($n=4$) and non-bipartite ($n=3$) lattices also manifests itself 
in conductance oscillations (see sections~\ref{sec:twoter}
and~\ref{sec:sq-tr}).
As shown in Ref.~\onlinecite{Browne84}, our
argument for $\phi_0/2$ periodicity in lattices with
odd elementary cycles also applies for the conductance
calculated using the Kubo formula.

\subsubsection{RG results}

The application of the strong disorder RG to
three and four-site clusters is discussed
in Appendix~\ref{app:warming}. In agreement with the preceding
argument we find a $\phi_0/2$ periodicity for the
three-site cluster with a symmetric distribution of $\epsilon_k$,
and a $\phi_0$ periodicity for the four-site cluster with 
$\epsilon_k=0$.
The permanent current is expanded according to
\begin{equation}
J(\phi)=\frac{\Omega_0}{\phi_0}
\sum_{n=1}^{+\infty} A_n \sin{\left(2 \pi n\frac{\phi}{\phi_0}\right)}
.
\end{equation}
The numerical results obtained by iterating the RG transformations
on the fractals with $n=3,4$, and on the Sierpinski gasket
(see Fig.~\ref{fig:RG-perma}), are
in agreement with the general argument in section~\ref{sec:general}.
The $\phi/\phi_0$ harmonics vanishes on the fractal with $n=3$
and on the Sierpinski gasket for a symmetric distribution of 
on-site energies. We obtain a strong $\phi_0$ harmonics for the
fractal with $n=4$.

It is interesting to compare these results with studies
of persistent current in mesoscopic rings. For a single
ring this current is expected to oscillate with a period
$\phi_0$\cite{Buttiker83}, and this prediction has been confirmed
in two experiments\cite{Chandrasekhar91,Mailly93}.
Upon ensemble-averaging with a fixed particle number,
the periodicity is expected to change to $\phi_0/2$
\cite{Bouchiat89,Oppen91,Altshuler91}, in agreement
with several observations on systems with a large number
of disconnected rings\cite{Levy90,Reulet95,Deblock02}.
Ensemble-averaging with a fixed chemical potential
instead of a fixed particle-number preserves a dominant
$\phi_0$ periodicity\cite{Cheung89}.

The argument of section \ref{sec:general}
suggests that this qualitative
difference between bipartite and non-bipartite lattices
regarding the behavior of the first harmonics of the 
permanent current and conductance
holds for a much larger class of systems. To check this,
we have also considered Euclidean lattices
in section \ref{sec:sq-tr}, namely the
square and the triangular lattices. 

\subsection{Two-terminal current}
\label{sec:twoter}

\begin{figure}
\includegraphics [width=.7 \linewidth]{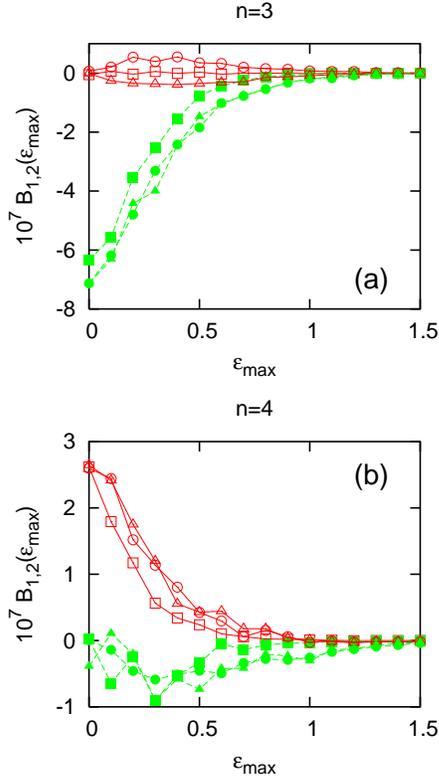}
\caption{Total current (in units of $e^2/h$)
evaluated at $\Gamma=0$,
averaged over $R$ between $R=15$ and $R=20$,
as a function of $f$, for the fractals with $n=3$
and $n=4$.
The different curves
within each panel correspond to the $\pm \epsilon$ model
with a symmetric distribution of $\epsilon_k$ ($\Box$
and $\blacksquare$), to the $+\epsilon$ model with
$\epsilon_k>0$ ($\circ$ and $\bullet$), and to the
$-\epsilon$ model with $\epsilon_k<0$ ($\triangle$,
$\blacktriangle$). The open symbols in red correspond to
the amplitude $B_1$ of the $\phi_0$ harmonics. The filled symbols
in green correspond to the amplitude $B_2$ of the $\phi_0/2$ harmonics.
Note the weak localization-like
positive magnetoconductance for the fractal with
$n=3$.
\label{fig:current}
}
\end{figure}

The evolution of the weight of the $\phi_0$
and $\phi_0/2$ harmonics as a function of 
$\epsilon_{\rm max}$
is shown in Fig.~\ref{fig:current}
for the fractals with $n=3$ and $n=4$. 
We focus here on the case $\Gamma=0$, corresponding
to a voltage equal to the maximal coupling.
The total current, obtained by evaluating the correlation
function ${\cal C}'_2(R)$ (see Eqs.~(\ref{eq:tr-form}) and
(\ref{eq:current-tot})), is expanded in a Fourier
series:
\begin{equation}
G(\phi)= \frac{e^2}{h} \sum_{n=0}^{+\infty} B_n
\cos{\left(2\pi n \frac{\phi}
{\phi_0} \right)}
.
\end{equation}
The variations of $B_1(\epsilon_{\rm max})$ and
$B_2(\epsilon_{\rm max})$ are shown on Fig.~\ref{fig:current}.
The two-terminal current
is almost $\phi_0/2$-periodic for
the fractal with $n=3$ and almost $\phi_0$-periodic for
the fractal with $n=4$, which is compatible with 
the previous behavior for the persistent current in section~\ref{sec:dia}.
The negative value of $B_2$ for the fractal with $n=3$ implies a positive
magnetoconductance at small flux, which looks like weak localization.
We deduce from the solution for a single triangular plaquette in
Appendix \ref{app:warming}, that the negative value of
$B_2$ for $\epsilon_{\rm max}=0$ for the fractal with $n=3$ is
due to a $t$-transformation followed by an $\epsilon$-transformation.
In this case the $\phi_0½$-periodicity originates from the interference
between the two time reversed paths around an elementary triangle, 
reminiscent of weak localization. We note also that because of infinite
disorder, the largest contribution to the energy comes from the first
RG transformations that couple to the magnetic flux. The permanent current
is thus dominated by the plaquettes on the smallest scale.

\subsection{Comparison with Euclidean lattices}

\label{sec:sq-tr}
\begin{figure}
\includegraphics [width=.9 \linewidth]{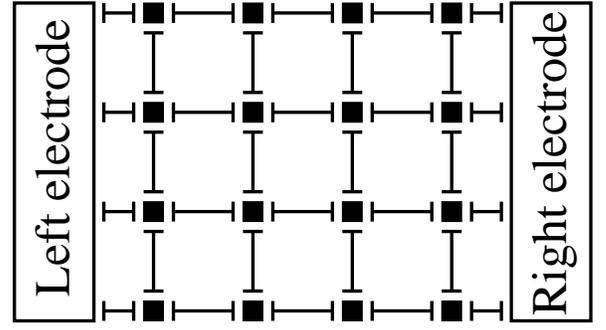}
\caption{Schematic representation of the $N\times N$
lattice of nodes 
and links used in the Green's function method
(with $N=4$). The lattice is
connected to a right and left electrode by extended contacts.
\label{fig:schema_Green}
}
\end{figure}

\begin{figure*}
\includegraphics [width=.8 \linewidth]{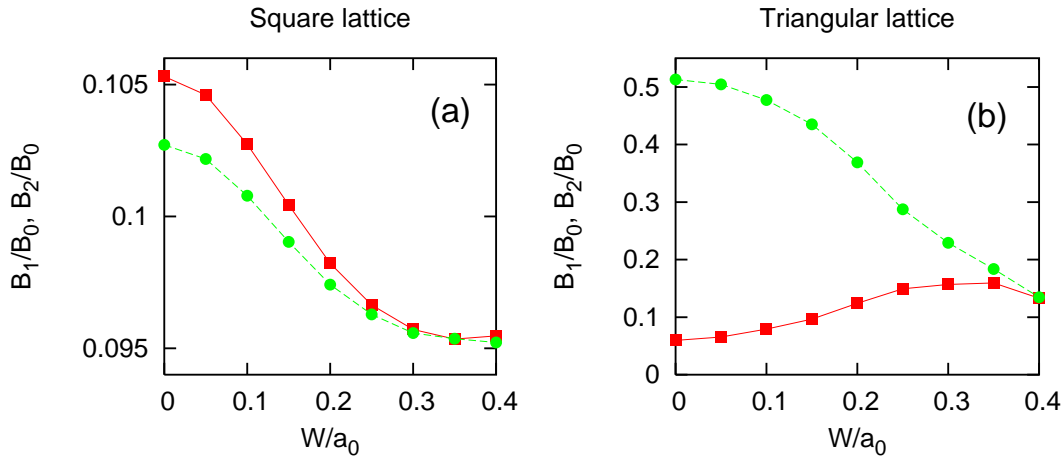}
\caption{Evolution of relative amplitude $B_1/B_0$
($\blacksquare$, red)
and $B_2/B_0$ ($\bullet$, green)
of the $\phi_0$ and $\phi_0/2$ harmonics
as a function of the strength of positional disorder $W/a_0$,
for the square (a) and triangular lattices (b). We use a lattice
of size $N \times N$, with $N=5$, and $k_F a_0=5$.
$B_2$ is positive for the triangular lattice,
unlike the case of the fractal with $n=3$ on Fig.~\ref{fig:current}.
\label{fig:sq-tr}
}
\end{figure*}

Using Green's function methods we calculate the conductance of an array of
nodes interconnected by single channel contacts (see
Fig.~\ref{fig:schema_Green}). 
Disorder is introduced in the coordinates $(x_k,y_k)$ of the nodes:
$x_k=x_k^{(0)}+\delta x_k$ and $y_k=y_k^{(0)}+\delta y_k$, 
where the $(x_k^{(0)},y_k^{(0)})$ correspond to the coordinates of the nodes
of the regular lattice having a lattice parameter $a_0$. The bare
Green's function depends on the distance between the nodes.
The variables
$\delta x_k$ and $\delta y_k$ are random, and chosen to be uniformly
distributed in the interval $[-W,W]$, with $W<a_0/2$.
A similar model has also been investigated in Ref.~\cite{Vidal}.
Some technical details regarding the Green's function treatment are
given in Appendix~\ref{app:Green}. 
The evolution of $A_1/A_0$ and $A_2/A_0$ as a function of
$W/a_0$ are shown on Fig.~\ref{fig:sq-tr} for the square and triangular
lattices. The triangular lattice is obtained by adding one diagonal on
the plaquettes of the square lattice. We obtain a strong $\phi_0/2$
harmonics for the triangular lattice while the two harmonics have 
approximately the same magnitude for the square lattice
(see Fig.~\ref{fig:sq-tr}).

For the tight-binding model on the
triangular lattice, the argument given in section~\ref{sec:dia}
can be adapted to prove that the Kubo conductivity is
$\phi_0/2$ periodic, when the on-site disorder has an even distribution,
and when the chemical potential vanishes. In our simulations there is
an extra phase $k_F R$ in the Green's function between two nodes
(see Eq.~\ref{eq:Green-3D}) so that the model that we consider is
not a tight-binding model. Nevertheless we find a behavior in a
qualitative agreement with the general argument
given in section~\ref{sec:general}: the
$\phi_0/2$ harmonics is small for the triangular lattice,
but not vanishingly small. We carried out a simulation with
$g_{k,l}^{(0)}=i \pi \rho_0$ instead of (\ref{eq:Green-3D}),
corresponding to a half-filled model
with $k_F R_{k,l}=\pi/2$, and found also a small $\phi_0$ harmonics
in this case. The amplitude of the $\phi_0/2$ harmonics is positive
for the triangular lattice, as opposed to the negative sign obtained
for the fractal with $n=3$ on Fig.~\ref{fig:current}. This indicates
that the weak localization-like behavior (associated to infinite disorder)
breaks down for the Euclidean, conventional disorder triangular lattice.
Similarly, the sign of the harmonics $A_2$ of the permanent current is
positive for the Sierpinski gasket (corresponding to conventional disorder)
while it is negative for the fractal with $n=3$ (corresponding to
infinite disorder). We thus see that the magnetoresistive effects
in infinite disorder systems are of a weak localization-like type
while a magnetoresistance opposite the one of weak localization
can be obtained for finite disorder systems.

The case of the square lattice is qualitatively compatible
with the results obtained by various groups on related models. 
In particular, quasi one-dimensional systems have been
studied by numerical\cite{Fourcade86,Avishai87} or analytical\cite{BD}
methods. A transition from $\phi_0$ to $\phi_0/2$ periodicity
has been found when disorder becomes large enough so that
the elastic mean free path is comparable to the lattice spacing.
Similar results for a two-dimensional square lattice have also
been reported\cite{Vidal}.
In the strong disorder regime, a very interesting 
$\phi_0$ to $\phi_0/2$ transition has been predicted by
Nguyen et al.\cite{Nguyen85a,Nguyen85b}, in a model where local random
potentials can take only two large and opposite values, with probabilities
$x$ and $1-x$. At small $x$, it exhibits
$\phi_0$-periodic oscillations in a Aharonov-Bohm geometry,
and $\phi_0/2$ appears for a concentration $x$ estimated 
around 0.05 for a square lattice. In all these systems,
the $\phi_0/2$ periodicity occurs when disorder is large enough
so that propagation paths interfering around a plaquette can be 
assumed to have a random relative phase.

\section{Conclusions}

In this paper disordered tight-binding models coupled to an external
magnetic field were studied on different fractal lattices. The fractals
have different type of topology (Sierpinski gasket and
fractals based on polygons with $n$ sides),
symmetry (bipartite and non bipartite) and fractal
dimension, varying between $d_f=1$ and $d_f=2$. We studied the
low-energy spectrum of the systems, the localization properties
(fraction of frozen sites), correlation functions and correlation
length and transport properties (permanent current and
two-terminal current). The main method of calculation is the numerical
implementation of a strong disorder RG approach. We generalized
the method to non-bipartite lattices and to
extrinsic diagonal disorder.  The results of numerical calculations
are analyzed by using exact arguments, phenomenological scaling
considerations and by comparing with results obtained by Green's
function methods on Euclidean lattices. To have a further comparison
and to obtain a classification of random fixed points in fractals we
have also studied the random antiferromagnetic Heisenberg model, as
well as the spin-glass model ({\it i.e.} with random ferro- and
antiferromagnetic couplings) on the same lattices and by the same type
of strong disorder RG method.

Generally infinite disorder fixed points are observed on the 
fractals based on polygons with $n$ sides,
both for the random antiferromagnetic Heisenberg model and
for the random tight-binding model without extrinsic diagonal
disorder. On the other hand a strong disorder ({\it i.e.} conventional
disorder) fixed point is found on the Sierpinski gasket and for the
spin-glass ($\pm J$) Heisenberg model for all types of fractals.  The
specific properties of low-energy fixed points realized in fractals
are found in some respect different from that already known in 1D
and 2D regular lattices.

For the random tight-binding model an interesting new feature of our study
is to consider non-bipartite lattices as well as extrinsic diagonal disorder. The
strong disorder RG method is shown to generate frozen sites (which do not
contribute to the average long-range correlations and to the conductance)
and to create a strongly interconnected effective cluster,
which, however is less
compact than that for $D \ge 2$ Euclidean lattices. The correlation function,
in the absence of diagonal disorder, is shown to exhibit a non-trivial algebraic decay,
whereas for extrinsic diagonal disorder there is a finite correlation length, the
value of which is related to the cut-off of diagonal disorder. The persistent
current is shown to exhibit a periodicity of $\phi_0/2$ or $\phi_0$, 
depending on whether
the elementary plaquettes in the lattices have an odd or even number of
sites, respectively.
The same type of periodicity is also observed in the two-terminal current.
Note that two distinct effects are involved: one is specific to bipartite
lattices for which only $t$-transformations are carried out. The other is
specific to a class of non bipartite lattices for which both $\epsilon$-
and $t$-transformations are carried out but the $\phi_0$ harmonics vanishes
for a symmetric distribution of on-site energies. Lattices with an odd number
of sites around each plaquette belong to this second class of non
bipartite lattices, that contains also other lattices such as the fractal
with $n=3$.

Finally we note that fractals are realized in different forms in 
nature\cite{mandelbrot}, for instance at
a percolation transition point. Our
results might have some relevance of understanding
the low-temperature dynamics and transport in these systems.

\section*{Acknowledgments}
The authors acknowledge many fruitful discussions with J.C. Angl\`es
d'Auriac. 
This work has been supported by the French-Hungarian cooperation program
Balaton (Minist\`ere des
Affaires Etrang\`eres - OM), the Hungarian National
Research Fund under  grant No OTKA TO34183, TO37323, TO48721,
MO45596 and M36803.

\appendix

\section{RG transformations of the tight-binding model}
\label{app:RG}

\begin{figure}
\includegraphics [width=1. \linewidth]{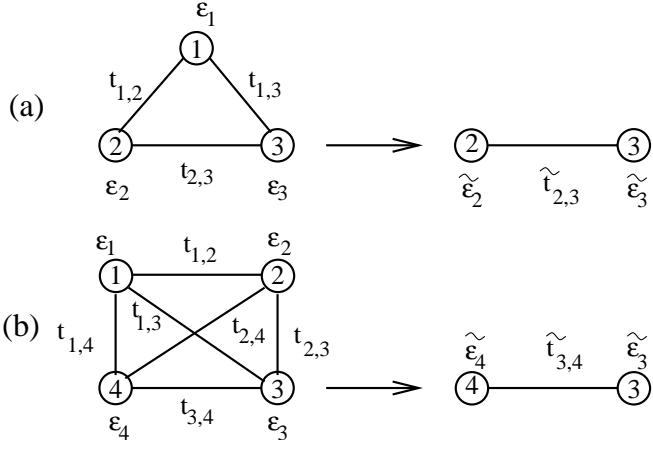}
\caption{(a) RG transformation in which site 1 is frozen.
(b) RG transformation in which the bond 1-2 is frozen.
\label{fig:RG-schema}
}
\end{figure}
In this Appendix we provide a short derivation of the
RG transformations of the random tight-binding model
on a general lattice, not necessarily bipartite.
The strongest gap can be due either to an on-site energy or
to a hopping energy. In the former case a fermion or a hole is
frozen at a given site. In the latter case a fermion
is frozen in a dimer.

\subsection{$\epsilon$-transformation}
Let us suppose that the strongest gap is due to the 
on-site energy $\epsilon_1$ of site
``1'' (see Fig.~\ref{fig:RG-schema}-(a)).
Considering first a two-site cluster $\{1,2\}$,
the renormalization of the on-site energies is given by
\begin{equation}
\tilde{\epsilon}_2 = \epsilon_2 - t_{2,1} \frac{1}{\epsilon_1} t_{1,2} 
,
\end{equation}
where $t_{1,2}$ abd $t_{2,1}$ are complex numbers such that
$t_{2,1}=(t_{1,2})^*$ (see Eq.~\ref{eq:t-def}).
Considering now a three-site cluster $\{1, 2, 3\}$, 
we obtain 
\begin{equation}
\tilde{t}_{2,3} = t_{2,3} -t_{2,1} \frac{1}{\epsilon_1} t_{1,3}
\label{eq:hopping-eps}
.
\end{equation}

\subsection{$t$-transformation}
Let us now suppose that the strongest gap is due to a hopping
integral $t_{1,2}$ between sites ``1'' and ``2'' and first
consider the three site clusters $\left\{1,2,3\right\}$.
The renormalization of $\epsilon_3$ is given by
\begin{equation}
\label{eq:epstilde-inter1}
\tilde{\epsilon}_3 =\epsilon_3
-t_{3,1} \frac{1}{t_{2,1}} t_{2,3}
-t_{3,2} \frac{1}{t_{1,2}} t_{1,3}
.
\end{equation}
Considering now
a cluster made of sites $\left\{1,2,3,4\right\}$,
the renormalization of the hopping amplitude is given
by 
\begin{equation}
\label{eq:hopping-t}
\tilde{t}_{3,4} = t_{3,4} 
-t_{3,1} \frac{1}{t_{2,1}} t_{2,4}
-t_{3,2} \frac{1}{t_{1,2}} t_{1,4}
.
\end{equation}

\subsection{Derivation of the RG equations}
As an example we provide a derivation of $t$-transformation
in a three-site cluster. The Hamiltonian is 
$h_0=t_{1,2} c_1^+ c_2+t_{2,1} c_2^+ c_1 + \epsilon_3
c_3^+ c_3$, and the perturbation $h'$ is made of the
$t_{1,3}$ and $t_{2,3}$ terms.
The unperturbed state is
\begin{equation}
|\psi_0\rangle = \frac{1}{\sqrt{2}}
\left[ \frac{t_{1,2}}{|t_{1,2}|} |1,0,1\rangle
-|0,1,1\rangle \right]
,
\end{equation}
where the basis is formed by the eight states
$|n_1,n_2,n_3\rangle$, with $n_1,n_2,n_3 \in \{0,1\}$.
The perturbed eigenstate takes the form $|\psi\rangle=
|\psi_0\rangle+P|\psi\rangle$, where $P$ is the projector in
the orthogonal to $|\psi_0\rangle$. The eigenstate can be expanded
according to 
\begin{equation}
|\psi\rangle=|\psi_0\rangle +
(E-h_0)^{-1}P h' |\psi_0\rangle + ...
,
\end{equation}
and the correction to the energy is given by
$\Delta E = \langle \psi_0|h'|\psi\rangle$.
After a straightforward calculation
we obtain
\begin{equation}
\Delta E = -\frac{|t_{1,3}|^2}{2|t_{1,2}|}
- \frac{|t_{2,3}|^2}{2 |t_{1,2}|}
-t_{3,1} \frac{1}{2 t_{2,1}} t_{2,3}
-t_{3,2} \frac{1}{2t_{1,2}}t_{1,3}
,
\end{equation}
where the last two terms contribute to the renormalization
of $\tilde{\epsilon}_3$ in Eq.~(\ref{eq:epstilde-inter1}).

\section{Transport formula}
\label{app:tr}
\begin{figure}
\includegraphics [width=.9 \linewidth]{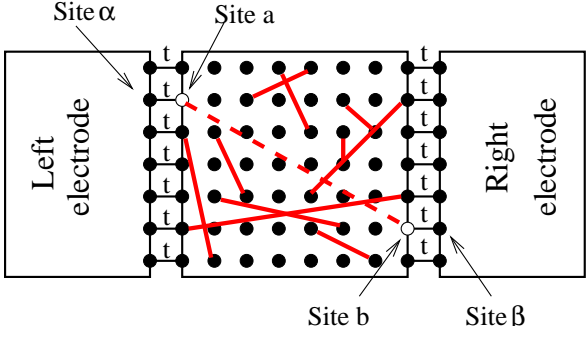}
\caption{Schematic representation of the system at energy
scale $\Omega$. Dimers of variable length (in red) are formed 
by the $t$-transformation.
The sites not involved in a dimer are either frozen by
an $\epsilon$-transformation or have not been yet decimated.
We have specialized two sites $a$ and $b$ in the random system,
and their corresponding sites $\alpha$ and $\beta$ in the
left and right electrodes. The sites $a$ and $b$ are
connected by a hopping $t_{a,b}$ (dashed line),
calculated from the RG.
We have represented the hopping
amplitudes $t$ making an extended contact between the
random system and the left and right electrodes.
For clarity we have represented
an Euclidian 2D random system but
we calculate the conductance in fractal
geometries in Fig.~\ref{fig:dessin}. 
$R$ is the distance between the left and right electrodes.
\label{fig:schema-tr}
}
\end{figure}

In this Appendix we provide a derivation of the transport
formula~(\ref{eq:tr-form}). At an energy scale $\Omega$,
the random system consists of ``dimers''
due to the $t$-transformations coexisting with
fermions frozen in a single site due to the
$\epsilon$-transformations.
Some of the dimers are frozen since they were coupled
by $|t_{k,l}|>\Omega$. Other dimers with $|t_{k,l}|=\Omega$
are being processed by the RG and the remaining dimers 
with $|t_{k,l}|<\Omega$
correspond to the ``active'' sites 

We consider two 3D normal electrode connected at the left and right
to the random system by two extended contacts
(see Fig.~\ref{fig:schema-tr}).
Two arbitrary sites $a$ and $b$ at the two
interfaces can be paired
by a hopping $t_{a,b}$ formed by a $t$-transformation.
We have $t_{a,b}=0$ if a bond $a$-$b$ has not been formed.
We note $t_{a,\alpha}$ and $t_{b,\beta}$ the hoppings
between sites $a$ and $\alpha$, and $b$ and $\beta$ (we
use $t_{a,\alpha}=t_{b,\beta}=t$. Within perturbation 
in $t_{a,\alpha}$ and $t_{b,\beta}$
the total differential conductance is given by
(i) the sum over all
pairs of sites $a$ and $b$ of the ``elastic cotunneling'' current
${\cal I}_{a,b}$ associated to the hoppings 
$t_{a,b}$, 
and (ii) the sum over all sites $a$ of the
``local'' currents ${\cal I}_{a,a}$. 
We expand
the elastic cotunneling current
${\cal I}_{a,b}$ according to
\begin{eqnarray}
\label{eq:Iab}
{\cal I}_{a,b} &=& 4 \pi^2 \frac{e}{h}
t_{a,\alpha}^2 t_{b,\beta}^2 
\rho_{\alpha,\alpha} \rho_{\beta,\beta}
\\\nonumber
&\times&
\int_{-\infty}^{+\infty}
 g_{a,b}^A(\omega) g_{b,a}^R(\omega)
\left[ f_R(\omega)-f_L(\omega) \right]
,
\end{eqnarray}
where $f_{R,L}(\omega)=f_T(e V_{R,L}-\omega)$
is the Fermi distribution function in the right
and left electrodes,
$\rho_{\alpha,\alpha}$ and $\rho_{\beta,\beta}$
are the density of state of the 3D left and right
electrodes, $g_{a,b}^A$ and $g_{b,a}^R$ are the advanced
and retarded propagators from $a$ to $b$ and $b$ to $a$,
$\omega$ is the energy. We suppose that voltages
$V_L=-V$ and $V_R=V$
are applied on the left and right electrodes.
We recognize in Eq.~(\ref{eq:Iab}) a term proportional
to
the integral over energy of
${\cal C}_2'(R)$ (see Eq.~\ref{eq:C2-def}).

There is however an additional ``local'' current:
\begin{equation}
\label{eq:Iaa}
{\cal I}_{a,a} = 4 \pi^2
\frac{e}{h}
t_{a,\alpha}^2
\rho_{\alpha,\alpha} 
\int_{-\infty}^{+\infty} \rho_{a,a}(\omega)
\left[f_{V_a,V_b}(\omega)-f_L(\omega) \right]
,
\end{equation}
where $f_{V_a,V_b}(\omega)$ is the out-of-equilibrium
distribution function in
the disordered system, that depends on the value of the voltages
$V_a$ and $V_b$. The distribution function $f_{V_a,V_b}(\omega)$,
determined in such a way as to verify current conservation,
is in general not equal to the Fermi distribution.
The local current vanishes if $V_b=0$ or $t_{b,\beta}=0$,
because in these cases the out-of-equilibrium distribution function
$f_{V_a,V_b}(\omega)$ is equal to $f_L(\omega)$.
Within the strong disorder RG hypothesis
the non percolating renormalized hoppings starting
at one interface and not ending at the other interface
do not contribute to the current
since the fermions cannot hop from one dimer to the
other, or from one dimer to one active site.
The density of states $\rho_{a,a}$ in Eq.~(\ref{eq:Iaa})
is thus evaluated only for the percolating bonds
frozen at energy $\Omega$ while ${\cal C}_2'(R)$ involves a 
summation over all percolating bonds that have been frozen 
at energy scales larger than $\Omega$
(see section~\ref{sec:calc-C}). 
The ``elastic
cotunneling'' term (\ref{eq:Iab}) proportional
to the correlation function
${\cal C}_2'(R)$ is thus much larger than the ``local''
term (\ref{eq:Iaa}) if $t=t_{a,\alpha}=t_{b,\beta}$ remains
finite while the area of the contacts is sent to infinity.
We have thus provided a derivation
of the transport formula given by Eq.~(\ref{eq:tr-form}).

\section{Tight-binding model on small clusters}
\label{app:warming}

Because of strong disorder the dominant contributions 
to the permanent current and conductance are due to the RG
transformations on the smallest length scale.
The periodicity of the permanent current can thus be
qualitatively understood with
simple networks made of a single plaquette.

\subsection{Triangular plaquette}

\subsubsection{$\epsilon$-transformation followed by a 
$t$-transformation}
Considering a triangular plaquette made of three
sites $\{1,2,3\}$ and pierced by a magnetic flux $\phi$,
we first suppose that $\epsilon_1$ is the strongest energy.
After eliminating the site ``1'', the renormalized couplings
are
given by
\begin{eqnarray}
\tilde{\epsilon}_2^{(1)}&=&\epsilon_2-t_{2,1}
\frac{1}{\epsilon_1} t_{1,2}\\
\tilde{\epsilon_3}^{(2)}&=&\epsilon_3-t_{3,1}
\frac{1}{\epsilon_1} t_{1,3}\\
\tilde{t}_{2,3}^{(1)} &=& t_{2,3}
-t_{2,1} \frac{1}{\epsilon_1} t_{1,3}
,
\end{eqnarray}
and the correction to the ground state energy after the first
transformation is
$\delta E^{(1)}=-|\epsilon_1|/2$. The correction to the ground
state energy due to the next $t$-transformation is
$\delta E^{(2)}=-|\tilde{t}_{2,3}^{(1)}|$, with
\begin{equation}
\left|\tilde{t}_{2,3}^{(1)}(\epsilon_1)\right|=
\sqrt{ A-\frac{B}{\epsilon_1}  \cos{(2\pi \phi / \phi_0)
}}
,
\end{equation}
with
$A = (t_{2,3}^{(0)})^2 +
(t_{1,2}^{(0)} t_{1,3}^{(0)}/
\epsilon_1)^2$ and 
$B = 2 t_{1,2}^{(0)} t_{1,3}^{(0)} t_{2,3}^{(0)},$
where $t_{i,j}^{(0)}$ corresponds to the value of the
hopping in the absence of magnetic flux.
Assuming a symmetric distribution $P(\epsilon_1)=P(-\epsilon_1)$,
the average ground
zero temperature state energy given by
\begin{equation}
U(\phi)=-\frac{|\epsilon_1|}{2}
-\int d\epsilon_1 P(\epsilon_1)
\left[ |\tilde{t}_{2,3}^{(1)}(\epsilon_1)|
+ |\tilde{t}_{2,3}^{(1)}(-\epsilon_1)| \right]
\end{equation}
is $\phi_0/2$-periodic.

\subsubsection{$t$-transformation followed by an $\epsilon$-transformation}
We suppose now that the sites $\{1,2\}$ are first frozen in a dimer.
The ground state energy is given by
\begin{equation}
E(\phi)=-t_{1,2}^{(0)}-\frac{1}{2} \left|
\epsilon_3- f(\phi/\phi_0) \right|
,
\end{equation}
with
\begin{equation}
f(\phi/\phi_0) = 2 \frac{t_{1,3}^{(0)}
t_{2,3}^{(0)}}{t_{1,2}^{(0)}} \cos{(2\pi \phi/\phi_0)}
.
\end{equation}
Assuming a symmetric distribution of on-site energies, we obtain
\begin{eqnarray}
U(\phi)&=&-t_{1,2}^{(0)}-\int_{|f(\phi/\phi_0)|}^{\epsilon_{\rm max}}
\epsilon_3 P(\epsilon_3) d\epsilon_3\\
&-&|f(\phi/\phi_0)|
\int_0^{|f(\phi/\phi_0)|} P(\epsilon_3) d\epsilon_3
.
\end{eqnarray}
The average ground state energy is $\phi_0/2$-periodic,
and the sign of the $\phi_0/2$ harmonics
is in agreement with Fig.~\ref{fig:RG-perma}-(a) for
$\epsilon_{\rm max}=0$.

\subsection{Square plaquette}
Only $t$-transformations
are performed for a bipartite lattice in the absence of extrinsic 
diagonal disorder. It is easy to show that two consecutive
$t$-transformations generate a $\phi_0$-periodic
ground state energy.
\section{Green's functions method}
\label{app:Green}

In this appendix we provide a brief description of the Green's function
method by which we calculate the conductance. We introduce nodes at the
vertices of a square lattice, interconnected by links (see
Fig.~\ref{fig:schema_Green}). We obtain a triangular lattice by adding
a link on one diagonal of the plaquettes. A node $n_k$ correspond to a Green's
function $g_{k,k}=i \pi \rho_0$ and a link between the nodes
$n_k$ and $n_l$
corresponds to a Green's function
\begin{equation}
g_{k,l}(\phi)=g_{k,l}^{(0)} \exp{\left(
\frac{2 i \pi}{\phi_0} \int_{{\bf r}_k}^{{\bf r}_l} {\bf A}.d {\bf r}
\right)}
,
\end{equation}
where ${\bf A}$ is the vector potential and $g_{k,l}(\phi)$ is the Green's
function in the absence of an applied magnetic flux. We choose
\begin{equation}
g_{k,l}^{(0)} =
\frac{\pi \rho_0}{k_F R_{k,l}} \exp{(i k_F R_{k,l})}
,
\label{eq:Green-3D}
\end{equation}
where $R_{k,l}$ is the distance between the nodes $n_k$ and $n_l$.
Eq.~(\ref{eq:Green-3D})
corresponds to the Green's function of a bulk metal with a Fermi
wave-vector $k_F$. We obtained similar results with other spatial
variations of the bare Green's functions.

The fully dressed Green' s function $G$
is obtained by interconnecting the nodes and links by
a hopping amplitude $T$ such that
$\pi T \rho_0=1$ corresponding to highly
transparent interfaces. The Green's function $G$ of the lattice of connected
nodes and links is obtained through the Dyson equation
$G = g + g \otimes \Sigma \otimes G$, where $\otimes$ is a short notation
for a convolution over time variables (becoming a simple product after
a Fourier transform to the frequencies), and a summation over the labels
of the graph of nodes and links. The symbol $\Sigma$ denotes the self-energy
corresponding to the connections between the nodes and links.
By using two times the Dyson equation we
obtain a linear set of equations by which we determine the Green's function
$G_{k,l}$ connecting the nodes $n_k$ and $n_l$.
The Kubo conductance corresponding
to two extended tunnel contacts at the left and right sides of the network
is then proportional to
\begin{equation}
{\cal G}(\phi) = \frac{e^2}{h} \sum_{k,l} | G_{k,l}(\phi) |^2
,
\end{equation}
where $k$ and $l$ run over all sites in the left-most and right-most
columns respectively.


\begin{thebibliography}{99}
\bibitem{note1}
The strongly reduced gaps for $n=3$, $f=0.25$ are due to the fact that
the number of sites of the lattice is odd and all transformations are
$t$-transformations except for the last transformation that is an $\epsilon$-transformation.
This last transformation leads to a much smaller gap, than the previous ones.
\end{thebibliography}
\end{document}